\documentclass[11pt,a4paper]{article}
\pdfoutput=1
\usepackage{jheppub}
\usepackage[utf8]{inputenc}
\usepackage{xcolor}
\usepackage{epstopdf}
\usepackage{graphicx}
\usepackage{epsfig}
\usepackage{dcolumn}  
\usepackage{bm}    
 \usepackage{caption}
 \usepackage{subcaption}
\usepackage{amssymb} 
\usepackage{amsmath,bm}
\usepackage{amsfonts}  
\usepackage{amsmath}  
\usepackage{slashed}  
\usepackage{enumitem}
\usepackage[mathscr]{euscript}
\usepackage{tabu}
\usepackage{epsfig}
\makeatletter
\g@addto@macro\bfseries{\boldmath}
\makeatother

\def\l1{{{1-loop}}}

\def\n1{\Bigg|_{n=1}}

\def\n{{(n)}}

\usepackage[T1]{fontenc} % if needed
\usepackage{tikz}
\usepackage{amsmath,amssymb}
\usepackage{relsize}
\usepackage{latexsym}
\usepackage{leftidx}
\usepackage{xcolor}
\usepackage{diagbox}
\usepackage[T1]{fontenc}
\usepackage{array}
\usepackage{makecell}
\usepackage{csquotes}
\usepackage{tikz}
\usepackage{enumitem}
\usepackage{setspace}
\usepackage{multirow}
\usepackage{amsmath,amssymb}
\usepackage{relsize}
\usepackage{latexsym}
\usepackage{leftidx}
\usepackage{xcolor}
\usepackage{csquotes}
\usepackage{tikz}
\usepackage{enumitem}
\usetikzlibrary{decorations.markings}
\usetikzlibrary{decorations.pathmorphing}
\usetikzlibrary{decorations.markings}
\usetikzlibrary{decorations.pathmorphing}
\usepackage{pifont}
\usepackage{hyperref}

\usepackage{bookmark}
  \title{\textbf{\textsf{Partition functions of $p$-forms from Harish-Chandra characters}}}
  \author{Justin R. David, Jyotirmoy Mukherjee}
\affiliation{\vspace{.1cm} Centre for High Energy Physics, \\ Indian Institute of Science,\\
C. V. Raman Avenue, Bangalore 560012, India.}
\emailAdd{justin@iisc.ac.in, jyotirmoym@iisc.ac.in}
\abstract{ 
We show that the determinant of the co-exact $p$-form  on spheres and anti-de Sitter spaces 
can be written as an integral transform of  
bulk and edge Harish-Chandra characters. 
The edge character of a  co-exact $p$-form contains 
characters of  anti-symmetric tensors of rank lower to  $p$  all the way to the zero-form. 
Using this result we evaluate the partition function of $p$-forms  and 
demonstrate that they 
obey  known properties under Hodge duality. 
We show that the partition function of conformal  forms in even $d+1$ dimensions, 
 on hyperbolic cylinders 
can be written as integral transforms involving only the  bulk characters. 
This  supports   earlier observations that entanglement entropy 
evaluated using partition functions on hyperbolic  cylinders  do not contain 
contributions  from the edge modes. For conformal coupled scalars we demonstrate that the
character integral representation of 
 the free energy on hyperbolic cylinders and branched spheres coincide. 
Finally we propose a  character integral representation for the 
 partition function of $p$-forms on branched spheres.
}
 
\begin{document}

\maketitle

\section{Introduction}

Partition functions of free fields 
 on spheres, anti-de Sitter spaces, hyperbolic cylinders, branched spheres, de Sitter spaces 
 are important ingredients for tests of  $AdS/CFT$ dualities, evaluation of entanglement entropies, 
 quantum corrections to black hole entropy and evaluation of anomaly coefficients in even dimensions 
  and the $F$-function in odd dimensions. 
 There are numerous works in the literature which highlight the importance of 
 these partition functions, a partial list of these references can be obtained 
 from the recent work of \cite{Anninos:2020hfj}. 
 It has always been useful to cast these one loop partition functions as 
 integrals over characters. One of the early instances  was in  \cite{David:2009xg}  where 
 the coincident heat kernel for arbitrary spin fields  on thermal $AdS_3$  was written as a transform of 
 the Harish-Chandra character of $SL(2, C)$. 
 In higher dimensional $AdS$ character integral representations were found in 
 \cite{Bae:2016rgm,Basile:2018zoy,Basile:2018acb} and 
 \cite{Beccaria:2014jxa,Skvortsov:2017ldz}, 
 in the  former series of works, the integral  also involved angular integrations over the additional Cartan directions. 
 
 Recently in \cite{Anninos:2020hfj} and \cite{Sun:2020ame} 
 a very useful  expression for the one loop partition function for 
 scalars, fermions and integer higher spin fields on spheres/euclidean patch of de Sitter space as 
 well as anti-de Sitter space was found.   This integral representation has only one integral over the 
Harish-Chandra character of the element  in $SO(1, 1)$ of these spaces just as the one found in \cite{David:2009xg}. 
 The expression  for the one loop path integral on the sphere  for higher spin fields on 
 $S^{d+1}$ can be summarised as follows 
 \begin{eqnarray}
 \log {\cal Z}  &=& \log {\cal Z}_G +  \log {\cal Z}_{{\rm chr} }, \\ \nonumber
\log {\cal Z}_{{\rm chr} } &=&   \int_0^\infty  \frac{dt}{2t}\frac{ 1+ e^{-t} }{ 1-  e^{-t} }
 \left( \chi_{{\rm bulk }} ( t) - \chi_{{\rm edge}} ( t) 
 \right) .
 \end{eqnarray}
 Here ${\cal Z}_G $ contains  the dependence of the  dimensionless 
 coupling constant of the theory.  The  dimensionless coupling is obtained 
 by considering  an appropriate   combination of the coupling  and the radius of the sphere. 
 ${\cal Z}_G$ also contains dependence of the volumes of the  gauge groups of the fields involved 
 in the one loop partition function. 
 The interesting contributions arise from  $ \log {\cal Z}_{{\rm chr} }$ which can be split into 2 parts. 
  The bulk contribution  can be written as 
 an integral transform of  Harish-Chandra characters of the group $SO(1, d+1) $, while the 
 edge term can be written as integral transform of Harish-Chandra character but with $d\rightarrow d-2$. 
  The contributions from the characters are interesting. 
 For example, in the case of even $d+1$, the coefficient of the logarithmic divergence which is 
 independent of regularization  and  can be extracted from the $1/t$ coefficient of the integrand in 
  $ \log {\cal Z}_{{\rm chr} }$. 
  In the case of odd $d+1$ one can extract the  IR finite  contribution to the partition function 
  by performing the integral using a regulator. 
  In  \cite{Anninos:2020hfj} expressions for  $ \log {\cal Z}_{{\rm chr} }$ were obtained for scalars, fermions 
and higher spin fields in arbitrary dimensions. 
This was then extended in \cite{Sun:2020ame}  to the one loop partition funciton 
for these fields in $AdS_{d+1}$ where now the integrand involves 
Harish-Chandra characters for  the group $SO( 2, d)$
 and  $SO(2, d-2)$ as bulk and edge contributions respectively. 

In this paper we obtain the character integral  representation of the one loop partition function for 
 $p$-forms on spheres
and anti-de Sitter spaces.   
It is known that after fixing gauge the path integral  
of $p$-forms  on spheres can be written as \cite{Obukhov:1982dt,Copeland:1984qk,Cappelli:2000fe}
\begin{align}\label{intro-pform}
    \mathcal{Z}_p[S^{d+1}]&=\Big[\frac{1}{\det_T \Delta_p}\frac{\det_T\Delta_{p-1}}{\det_T\Delta_{p-2}}\cdots\big(\frac{\det_T\Delta_1}{\det'\Delta_0}\rm{Vol} \; S^{d+1} \big)^{(-1)^p}\Big] ^{\frac{1}{2}},
\end{align}
where $\Delta_p$ is the   Hodge-de Rham operator acting on a  form of degree $p$ on the sphere.
The subscript $T$ in the determinant refers to the fact that the determinant is taken 
over co-exact or transverse $p$-forms.  The prime in the determinant of the $0$-form 
refers to the fact that it does not include the zero mode. 
Finally the ${\rm Vol} \;S^{d+1}$ is the volume of the $d+1$ dimensional sphere of radius 
$R$. 
From  (\ref{intro-pform}) we see that the key ingredient in the one loop partition function 
of the $p$-form is the determinant of the co-exact $p$-form. 
Starting from the eigen values of the  Hodge-de Rham 
Laplacian of co-exact $p$-forms on $S^{d+1}$ and their 
degeneracies, 
we show that the character part of the determinant of the co-exact $p$ form 
\begin{eqnarray}\label{result-1}
-\frac{1}{2} \log ( {\det}_T \Delta_p^{S^{d+1}}   )  = 
\int_0^\infty \frac{dt}{2t} \frac{ 1+ e^{-t}}{ 1 - e^{-t}} 
  \left(  \sum_{i = 0}^p ( -1)^i  \chi^{dS}_{( d- 2i,  p - i )} (t)  \right), 
\end{eqnarray}
where $ \chi^{dS}_{( d,  p )}(t)  $ is the $SO( 1, d+1)$  Harish-Chandra character of the anti-symmetric 
tensor of rank $p$, $\Delta = \frac{d}{2}  + i \nu$ representation with $ i \nu = \frac{d}{2} - p $.
This is  given by 
\begin{equation} \label{dschr}
 \chi^{dS}_{( d,  p )}( t)   = {d\choose p} \frac{ e^{ - ( d- p) t }  + e^{ - p t} }{ ( 1- e^{- t} )^ d}  .
 \end{equation}
 Form (\ref{result-1}) we see the bulk  and 
 the edge characters  on the sphere  $S^{d+1} $ for the co-exact  $p$-form is given by 
 \begin{equation}
\chi_{{\rm bulk}, \, ( d, p )   }^{dS}( t)   = \chi^{dS}_{( d, p )}(t)  , \qquad
\chi_{{ \rm edge\, {( d, p )} } }^{dS}(t)  =  -   \sum_{i = 1}^p ( -1)^i  \chi^{dS}_{( d- 2i,  p - i ) }(t)   .
\end{equation}
Note that the edge character involves characters in lower dimensions. 
The dimensions are lowered by units of $2$, while the degree of the $p$-forms are lowered 
by units of $1$ all the way to $p =0$. 
It is clear that if $2p>d$, the character involves powers of $e^{+t}$, therefore these are the 
`naive' characters in the sense of \cite{Anninos:2020hfj} and should be replaced by 
the  corresponding `flipped' character. 
Thus the character representation of the determinant of the co-exact $p$-form can be written 
as difference of the bulk and edge character. 
Using this representation of the determinant of the co-exact $p$-form, we  evaluate the 
path integral of the $p$ forms on the spheres $S^{d+1}$ for $ 2\leq d \leq 13$. 
It is clear from (\ref{intro-pform}) and (\ref{result-1}) that the partition function 
of the $p$-form is a difference of the bulk and the edge contribution. 
To emphasise this again, note that the path integral involves a product of determinants 
of co-exact forms from $0$ to $p$ (\ref{intro-pform}).
Therefore to evaluate the free energy of the $p$-form 
 we would need to  perform a sum of  (\ref{result-1}) from $p=0$ to $p$ with alternating signs.

The integral in (\ref{result-1})  can be regularised using the methods
of \cite{Anninos:2020hfj}.  
For even $d+1$ we show that the coefficient of the log divergence of the partition function  which is 
renormalization group invariant  can be extracted easily from the coefficient of 
the $t^{-1}$ term in the small $t$ expansion of the integrand. 
This coefficient 
precisely agrees with  earlier results in literature. 
For odd $d+1$ we regulate the integral to obtain 
the infrared finite part of the partition function and again the results agree with previous 
results in literature.
For both even and odd dimensions we observe  that these results obey known properties 
under Hodge duality. 

We then repeat the analysis for $p$-forms in anti-de Sitter space of $d+1$ dimensions. 
Here again the key ingredient to evaluate the partition function is 
the  determinant of the co-exact $p$ form  on $AdS_{d+1}$. 
Starting from the Plancherel measure and the eigen values of the Hodge-deRham Laplacian,
we show that the character part of  determinant  of  co-exact $p$ form on $AdS_{d+1}$  for even $d+1$ is 
given by 
\begin{eqnarray}
-\frac{1}{2} \log ( {\det}_{T}\Delta_p^{AdS_{d+1}}  )  = \int_0^\infty \frac{dt}{2t} 
\frac{ 1+ e^{-t}}{ 1- e^{-t}} \left ( \chi_{\rm bulk}^{AdS}( t)  -  \chi_{\rm edge}^{AdS}  ( t) \right) 
, \qquad  d+1 \; \hbox{even}  
\end{eqnarray}
where
\begin{eqnarray} \label{adschr}
\chi_{\rm bulk}^{AdS} (t) = \chi^{AdS}_{(d, p) } ( t) ,  \qquad
\chi_{\rm edge}^{AdS} (t)  = - \sum_{i =1}^p  (-1)^i \chi^{AdS}_{(d- 2i , p-i )  } ( t) .\\ \nonumber
\end{eqnarray}
$\chi^{ AdS}_{d, p }(t)  $ is the  $SO(2, d) $ Harish-Chandra character of the 
anti-symmetric tensor of rank $p$, $\Delta  = \frac{d}{2} + i \nu$ representation 
with $i \nu = \frac{d}{2} - p $ which is given by 
\begin{equation}
 \chi^{AdS}_{( d,  p )}( t)   = {d\choose p} \frac{ e^{ - ( d- p) t }   }{ ( 1- e^{- t} )^ d}  .
 \end{equation}
 Similarly for $AdS_{d+1}$ with $d+1$ odd is given by 
 \begin{eqnarray}
-\frac{1}{2} \log ( {\det}_{T}\Delta_p^{AdS_{d+1}}  )  =  \frac{  \log(\tilde R) }{2\pi i} \oint_C\frac{dt}{2t} 
\frac{ 1+ e^{-t}}{ 1- e^{-t}} \left ( \chi_{\rm bulk}^{AdS}( t)  - \chi_{\rm bulk}^{AdS}( t)  \right) 
, \qquad  d+1 \;  \hbox{odd} . \nonumber \\
\end{eqnarray} 
Here the contour $C$ is a small circle around the origin, and $\tilde R$ is the dimensionless 
IR cutoff. It can be taken as the ratio of the radial cutoff on $AdS$  to the radius 
of $AdS$.   The characters $\chi_{\rm bulk}^{AdS}( t)  , \chi_{\rm bulk}^{AdS}( t)  $ are defined in (\ref{adschr}). 
We perform the following consistency check on our results for the  partition function of $p$-forms:
One  loop free energies 
  on even dimensional spaces are determined by the 
trace anomaly of the theory  leads to the prediction that ratio of free energies in $AdS_{d+1} $ to 
$S^{d+1}$ is given by 
\begin{eqnarray} \label{testadss}
\frac{ \log {\cal Z}_p [AdS_{d+1}] }{
\log {\cal Z}_p[S^{d+1}]  }
=  \frac{1}{2} \qquad \hbox{for even} \quad d+1.
\end{eqnarray}
We verify that this relation is satisfied by the character integral representation of the 
partition function.

We then study conformal invariant  fields on hyperbolic cylinders. The fact that hyperbolic cylinders 
can be conformally mapped to spheres suggests that partition function of these fields on these spaces 
should be identical. 
Indeed in \cite{Klebanov:2011uf} it was verified through direct calculations that partition functions of 
conformal scalars and fermions  in $S^1\times AdS_2 $ precisely agrees with that on $S^3$. 
More recently in \cite{Nishioka:2021uef}, the free energies  of conformal scalars on  hyperbolic cylinders 
and spheres were shown to agree to  $d=100$ by explicit calculations
\footnote{See statement around   equation 4.52 of 
\cite{Nishioka:2021uef}.}.
Since we have found that partition functions are integral transforms of characters, we can ask
that if the integrands that occur in the character representation of the partition functions on hyperbolic cylinders
agree with the integrands of the partition functions on spheres. 
We find that indeed that for conformal scalars we  can sum over all the Kaluza-Klein 
modes on the $S^1$ of hyperbolic cylinders and show that the character which to begin with 
was an $AdS$ character becomes the character on the sphere.  The character integral representations
on spheres and hyperbolic cylinders coincide and therefore provides a proof that the 
free energies of conformal scalars on these spaces are same. 

Conformal $p$-forms occur in even $d+1$ dimensions. 
We show that  the logarithmic divergence 
of the gauge invariant partition function of the conformal $\frac{d-1}{2} $-form 
in $S^1 \times AdS_{d}$  can be written as the following transform of characters. 
\begin{eqnarray}
\log{\cal Z}_{\frac{d-1}{2}} [S^1 \times AdS_{d} ]
&=& \frac{ \log (\tilde R) }{2\pi i }  
\int_{C} \frac{dt}{2t} \frac{ 1+ e^{-t}}{ 1- e^{-t}} 
\sum_{i = 0}^{\frac{d-1}{2} } ( -1)^i \chi^{dS}_{(d, \frac{d-1}{2} - i) } (t), 
\qquad d+1  \; \hbox{even}  \nonumber \\
\end{eqnarray}
where $\chi^{[dS]}_{(d, p)}$  is given in (\ref{dschr}). Note the the integral 
contains $dS$ characters associated with the sphere $S^{d+1}$, it does not 
contain edge characters. In fact the integrand is precisely the bulk character of 
the conformal form on the sphere $S^{d+1}$. This is consistent with earlier observations 
found by studying entanglement entropy \cite{Huang:2014pfa,Donnelly:2014fua,Donnelly:2015hxa,Nian:2015xky,Dowker:2017flz,David:2020mls}. 
Entanglement entropy of theories with gauge symmetries  
evaluated  using partition functions on hyperbolic cylinders contain only the bulk contributions 
and misses out on the edge terms.

Finally  we study the co-exact $1$-form on the branched sphere $S_q^{d+1}$ with branching $q$. 
Using the known spectrum we write its free energy in terms of characters. 
From this result we are led to propose that the free energy of 
the co-exact $p$-form on branched spheres is given by 
\begin{eqnarray} \label{branched}
-\frac{1}{2} {\rm det}_T ( \Delta_p^{S_q^{d+1}} ) =
 \int_0^\infty \frac{dt}{2t} \left\{ \left( \frac{ 1 + e^{-  \frac{t}{q}  }}{ 1- e^{-\frac{t}{q} } } \right) 
 \chi_{(d, p)}^{dS} (t)  + 
  \left( \frac{ 1 + e^{-  t  }}{ 1- e^{-t} } \right) 
\sum_{i =1}^p  (-1)^i  \chi_{( d- 2i , p )}^{dS} (t)  \right\} . \nonumber \\
\end{eqnarray}
Note that the branching affects only the kinematic factor in front of the bulk character.  
The kinematic factor in front the edge character remains invariant under branching. 
We verify the proposal in  (\ref{branched}) by evaluating the partition function 
$p$-forms on branched spheres 
using (\ref{intro-pform}) and comparing to existing results in the literature. 

The organization of the paper is as follows: In section \ref{section1} we  write the free energy of 
$p$-forms on spheres in terms of characters, evaluate trace anomaly coefficients 
for even dimensional spheres and $F$-terms for odd spheres. 
We also demonstrate that these partition functions satisfy Hodge duality properties 
known in literature. 
In section  \ref{section2} we examine $p$-forms on $AdS$ spaces. 
One of important steps in evaluating the $p$-form Free energy  as 
a transform of Harish-Chandra characters is to to construct 
the Fourier transform of the Plancherel measure of the co-exact $p$-forms in 
$AdS_{d+1}$ 
We do this in section \ref{detco2}. We also verify the prediction (\ref{testadss}). 
Finally in section \ref{section3} we obtain character integral representations for 
free energies of conformal $p$-forms including the conformal scalar on 
hyperbolic cylinders. In the section \ref{branch} we determine the character integral 
representation  for free energies of $p$-forms on branched sphere by 
extrapolating the result for the $1$-form. 
The appendices contain a list of Harish-Chandra characters used in the paper
and an alternate approach to evaluate the Fourier transform of the 
Plancherel measure.

\section{$p$-forms on  spheres} \label{section1}

The partition function of gauge fixed $p$-form field on a sphere in total $d+1$  dimension is given by
\cite{Obukhov:1982dt,Copeland:1984qk,Cappelli:2000fe}
\begin{align} \label{gfpform}
    \mathcal{Z}_p[S^{d+1}]
    &=\Big[\frac{1}{\det_T \Delta_p}\frac{\det_T\Delta_{p-1}}{\det_T\Delta_{p-2}}\cdots\big(\frac{\det_T\Delta_1}{\det'\Delta_0}{\rm Vol }S^{d+1}\big)^{(-1)^p}\Big]^{\frac{1}{2}}.
\end{align}
$\det_T\Delta_p$ denotes the determinant of Hodge de-Rham Laplacian of the co-exact $p$-forms.
The prime in 
$\det'\Delta_0$  refers to the determinant of $0$-form or scalar without the zero modes. 
${\rm Vol}  S^{d+1}$ refers to the volume of the $d+1$ dimensional sphere. 
This arises due to the fact that the scalar has a zero mode, and integration over this zero mode 
results in the volume. 
Therefore the building block of the partition function of  $p$-forms is the determinant of the 
co-exact $p$ form. 
In section \ref{sectiondetcop} we will show that upon choosing an appropriate regulator, we can write the 
determinant of the co-exact $p$-form in terms of Harish-Chandra characters. 
Then in section (\ref{sectionlogdiv}) 
 use these determinants and evaluate the coefficient of the logarithmic 
divergence in partition function of $p$-forms on  even dimensional spheres. We will demonstrate that the result obeys 
known properties under Hodge duality. 
In section  (\ref{sectionfterm}) 
we evaluate the infrared finite term in the partition function for  $p$-forms on odd dimensional spheres and observe that they satisfy Hodge duality. 

\subsection{Determinant of co-exact $p$-forms as character integrals}
\label{sectiondetcop}

From the definition of the determinant in terms of its eigen values we have 
\begin{eqnarray}\label{detegi}
-\frac{1}{2} \log ( {\rm det}_T \Delta_p^{S^{d+1}} )  = - \sum_{n=1}^\infty 
\frac{1}{2} g_{n}^{(p)} \log ( \lambda_{n }^{(p)} ) ,
\end{eqnarray}
where the eigen values $\lambda_{n}^{(p)}$ and the degeneracies $g_{n}^{(p)}$ of the Hodge-deRham Laplacian 
of  co-exact $p$-forms on $S^{d+1}$ are given by  \cite{Copeland:1984qk}
\begin{equation} \label{degeig}
    \begin{split}
        \lambda_{n}^{(p)}&=(n+p)(n-p+d),\\
             g_{n}^{(p)} &=\frac{(2n+d)\Gamma[n+d+1]}{\Gamma[p+1]\Gamma[d-p+1]\Gamma[n](n+p)(n+d-p)}.
    \end{split}
\end{equation}
In (\ref{detegi})  the summation is from  $n=1$ to infinity for all values of $p$.  Note that this includes
the case of $p=0$ for which the zero eigen value is not part of the gauge fixed determinant.
Here we have focussed on the part of the one loop determinant that is independent of the radius of 
the sphere as well as the coupling of the theory. 
We now  replace the logarithm by the identity
\begin{equation}\label{logiden}
-\log y  = \int_0^\infty \frac{d\tau}{\tau} ( e^{-y \tau} - e^{-\tau} ) .
\end{equation}
Substituting this identity in (\ref{detegi}), we obtain 
\begin{eqnarray}\label{detegi2}
-\frac{1}{2} \log ( {\rm det}_T \Delta_p^{S^{d+1}} )  = \int_0^\infty \frac{d\tau}{2\tau} 
\left( \sum_{n=1}^\infty  g_{ n}^{(p)} ( e^{-\tau \lambda_{n}^{(p)}  } -   e^{-\tau})  \right) .
\end{eqnarray}
Here we wish to point out that our treatment differs from that of \cite{Anninos:2020hfj} in which the
identity (\ref{logiden}) was not used and therefore the second term in (\ref{detegi})  was 
not present for \cite{Anninos:2020hfj}. 
As such the integral in (\ref{detegi2}) is convergent at $\tau=0$ provided the 
second term can be regularized. 
To regularize the  second term we follow the approach introduced by \cite{Giombi:2015haa}. 
First note that the large $n$ behaviour of the degeneracies is given by 
\begin{align}
    g_n^{(p)}&=\left(\frac{1}{n}\right)^{-d}
     \left(\frac{2}{\Gamma (d-p+1) \Gamma (p+1) }+O\Big(\frac{1}{n} \Big) 
    \right).
\end{align} 
Therefore the 
 sum over the degeneracies can be performed by  first choosing $d$ to be a sufficiently negative 
 and continuing this result to positive values of $d$. We will refer to this as 
 `dimensional regularization'.
The result is given by  \cite{Giombi:2015haa,Raj:2016zjp}
\begin{eqnarray} \label{sumdegen}
\sum_{n=1}^\infty  g_{ n}^{(p)}  = - \cos p\pi  =  (-1)^{p+1}.
\end{eqnarray}
Using this result in (\ref{detegi2})  we obtain 
\begin{eqnarray}
- \frac{1}{2} \log ( {\rm det}_T \Delta_p^{S^{d+1}} )   = \int_{0}^{\infty}\frac{d\tau}{2\tau}
 \left(\sum_{n=1}^{\infty}g_{n}^{(p)} e^{-\tau\lambda_{n}^{(p)}  } +(-1)^p e^{-\tau} \right).
     \end{eqnarray}
     Now that we  have regulated the $\tau=0$ limit using dimensional regularization of the 
     sum there is no need of introducing a UV regulator as in \cite{Anninos:2020hfj}.
     However we find it convenient to introduce an  $\epsilon $ regulator.
     This will help us keep track of the branch cuts 
     in the $\tau$-plane that arise in the integrand and will not serve as a UV regulator. 
      Indeed finally we will take the $\epsilon\rightarrow 0$ limit. 
     We will also indicate how the branch cuts are present if $\epsilon$ was not introduced. 
    This results in 
    \begin{equation}\label{detegi3}
- \frac{1}{2} \log ( {\rm det}_T \Delta_p^{S^{d+1}} )    =\int_{0}^{\infty}\frac{d\tau}{2\tau}e^{-\frac{\epsilon^2}{4\tau}}\left(\sum_{n=1}^{\infty}g_{n}^{(p)} e^{-\tau\lambda_{n}^{(p)} }+(-1)^p \right) .
    \end{equation}
    For the second term  $\epsilon$ is introduces by the change of variables
     $\tau \rightarrow \frac{\epsilon^2}{4\tau}$.
    We will soon see that this choice leads us to the character integral representation for the 
    one loop partition function.  What we will obtain is the `naive' character in the sense of 
    \cite{Anninos:2020hfj}. 
     The integral is not regulated in the IR yet.  To extract IR finite terms 
    we need  another regulator as in  \cite{Anninos:2020hfj}. In section \ref{sectionfterm}
    we will  discuss the procedure to  extract the IR finite terms. 
    Let us go back to the integral in (\ref{detegi3}), 
by  analytically continuing in $n$, from (\ref{degeig}) we  notice that 
   \begin{eqnarray}\label{degproperty}
  && g_{n}^{(p)}  =0 , \qquad  \hbox{for} \qquad  -( p-1) < n < 0 ,  \\ \nonumber
     && g_{ -p}^{(p)} =(-1)^p , \qquad \hbox{and} \qquad     \lambda_{-p }^{(p)} = 0.
    \end{eqnarray}
   Using these properties of the degeneracies and eigen values, we can continue the sum in 
   \eqref{detegi2} to continue to $n =-p$ for the co-exact $p$-form.  This 
   extension resulted naturally  due to  the properties of the degeneracies 
    given in \eqref{degproperty}, which as far as we are aware has not been noted earlier. 
   We obtain
   \begin{equation} \label{beginstep}
  -  \frac{1}{2} \log ( {\rm det}_T \Delta_p^{S^{d+1}} )  
   =\int_{0}^{\infty}\frac{d\tau}{2\tau}e^{-\frac{\epsilon^2}{4\tau}}
   \left(\sum_{n=-p}^{\infty}g_{n}^{(p)} e^{-\tau\lambda_{n}^{(p)}}\right).
   \end{equation}
   Such an analytical continuation of the sums were also seen for higher spin fields 
   by \cite{Anninos:2020hfj} using a different approach. 
   As we have mentioned our starting point (\ref{detegi2}) and the regulator used 
   is different from that of \cite{Anninos:2020hfj}. To show our approach 
   yields the same results we have repeated the analysis 
   for partition function of massless symmetric rank $s$ tensor using our approach 
   in appendix \ref{spin-s-sum}
    and obtained the same conclusions of (\cite{Anninos:2020hfj}).
    
    We now follow the steps of  \cite{Anninos:2020hfj} and carry out the sum over $n$, 
  After writing $\lambda_{n}^{(p)} $ as difference of squares we get
  \begin{equation}
  \frac{1}{2} \log ( {\rm det}_T \Delta_p^{S^{d+1} })  =
      \int_{0}^{\infty}\frac{d\tau}{2\tau}e^{-\frac{\epsilon^2}{4\tau}}\left(\sum_{n=-p}^{\infty}g_{n}^{(p)}
      e^{-\tau(n+\frac{d}{2})^2}e^{-\tau\nu_p^2}\right),
  \end{equation}
  where $\nu_p^2=-(p-\frac{d}{2})^2$. 
  We can now linearise the sum over $n$ by using the  Hubbard-Stratonovich  trick. 
  \begin{equation} \label{contint}
   \sum_{n=-p}^{\infty}g_{n}^{(p)} e^{-\tau(n+\frac{d}{2})^2}
   =\int_{C} \frac{du}{\sqrt{4\pi\tau}}e^{-\frac{u^2}{4\tau}}f_p(u).
\end{equation}
Here the contour $C$ runs from $-\infty $ to $\infty$ slightly above the real axis  as in figure \ref{fig1}. 
and 
\begin{equation}
f_p(u) =\sum_{n=-p}^{\infty}g_{n}^{(p)} e^{i u(n+\frac{d}{2})}.
\end{equation}
Substituting the  degeneracies $g_{n}^{(p)}$ given in (\ref{degeig}) this sum can be performed resulting in 
\begin{eqnarray} \label{sumfpu}
&& f_p(u) =\frac{e^{\frac{1}{2} i (d+2) u} \Gamma (d+3)}{(d-2 p) \Gamma (d-p+1)}\nonumber \\ \nonumber 
&& \times \bigg[ \frac{\left(e^{i u}\right)^{-d+p-1} \left(B(e^{iu};d-p+1,-d-1)-2 B(e^{iu};d-p+1,-d-2)\right)}{\Gamma (p+1)}\nonumber\\  \nonumber
    & & +2 \,[ _2\tilde{F}_1\left(d+3,p+1;p+2;e^{i u}\right) ]-\, _2\tilde{F}_1\left(d+2,p+1;p+2;e^{i u}\right)\bigg] 
    \\ 
  &&  \qquad\qquad\qquad + (-1)^p 
    e^{i u \left(\frac{d}{2}-p\right)},
\end{eqnarray}
where $B(e^{iu};d-p+1,-d-1)$ is the incomplete beta function  defined 
as
\begin{equation}\label{defbeta}
B(z, a, b) = \int_0^z  t^{a-1} ( 1- t)^{b-1} \, dt
\end{equation}
and ${}_2\tilde{F}_1(a,b,c;z)$ is the regularised hypergeometric function.
\begin{align}\label{defhyper}
    _2\tilde{F}_1(a,b,c;z)=\frac{_2F_1(a,b,c;z)}{\Gamma[c]}.
\end{align}
The last line of $f_p(u)$  in (\ref{sumfpu}) comes from term $n=-p$ in the sum. We can consider this term as 
the contribution from the zero mode. 
Substituting this sum we obtain 
\begin{equation}\label{sphersuman1}
-\frac{1}{2} \log ( {\rm det}_T \Delta_p^{S^{d+1} })  = \int_0^\infty  \frac{d\tau}{2\tau}  \int_C \frac{du}{ \sqrt {4\pi \tau}}
e^{ - \frac{\epsilon^2 + u^2 }{4\tau}}
e^{\tau \nu_p^2} f_p( u ) .
\end{equation}
Let us now perform the $\tau$ integral  which results in 
\begin{equation}
-\frac{1}{2} \log ( {\rm det}_T \Delta_p^{S^{d+1} })   =
\int_{C}\frac{du}{ 2\sqrt{u^2+\epsilon^2}}\left(e^{-\nu_p\sqrt{u^2+\epsilon^2}}f_p(u)\right).
\end{equation}
At this stage it is good to point out that all the previous steps could have also been performed 
with $\epsilon=0$. However in the above step one would have obtained $e^{- \nu_p|u|}/|u|$ indicating 
the presence of branch cut. If one worked with $\epsilon=0$ one needs to  keep track of this, it is easier 
to do this with $\epsilon \neq 0$, therefore we continue as before. 
We  deform the contour 
$C$ from the real line to the contour  $C'$ which runs on the both sides of the branch cut on the imaginary axis 
originating at $u = i \epsilon$ on the  $u$-plane as shown in figure \ref{fig1}. 
Substituting $u = i t$ we obtain
\begin{equation} \label{endstep}
-\frac{1}{2} \log ( {\rm det}_T \Delta_p^{S_{d+1} })   =
\int_{\epsilon}^{\infty}\frac{dt}{2\sqrt{t^2-\epsilon^2}}\left(e^{i\nu_p\sqrt{t^2-\epsilon^2}}+e^{-i\nu_p\sqrt{t^2-\epsilon^2}}\right)f_p(i t).
\end{equation}
We can now take $\epsilon\rightarrow 0$ and the resulting integral is on the positive real axis in the $t$- plane
as shown in figure \ref{fig2}. 
Now it can be shown by  explicit check that the following remarkable identity holds
 \footnote{We have checked this for all values of $p, d\leq 14$. }
\begin{eqnarray}
( e^ { ( \frac{d}{2} - p) t } + e^{  -( \frac{ d}{2} - p ) t} ) f_p( i t) 
&=& \frac{ 1+ e^{-t} }{ 1- e^{-t} } \sum_{i =0}^p ( -1)^i \chi^{dS}_{( d- 2i , p -i ) } ( t) , \\
\label{defchr}
\chi^{dS}_{(d,p)}(t) &=&\binom{d}{p}\frac{e^{-t(d-p)}+e^{-tp}}{(1-e^{-t})^{d}}.
\end{eqnarray}
Therefore the partition function of the co-exact $p$-form reduces to 
 \begin{align}\label{chrintrep}
  -\frac{1}{2} \log ( {\rm det}_T \Delta_p^{S^{d+1} })  
  &=\int_{0}^{\infty}\frac{dt}{2t}\frac{ 1+ e^{-t}}{ 1- e^{-t} }
  \left(\sum_{i=0}^{p}(-1)^i \chi^{dS}_{(d-2i,p-i)}(t)\right) 
\end{align}
which is an integral representation in terms of $SO(1, d+1 -2i )$ Harish-Chandra characters. 
The term $i =0$, is the `naive' bulk character  in the sense of \cite{Anninos:2020hfj}
of the co-exact $p$-form, while all terms $i\geq 1$ constitute the `naive' edge characters.

Just as a check, let us examine the co-exact $1$-form, we obtain 
\begin{eqnarray}
  -\frac{1}{2} \log ( {\rm det}_T \Delta_1^{S^{d+1} })   &=&
  \int_{0}^{\infty}\frac{dt}{2t}\frac{ 1+ e^{-t}}{ 1- e^{-t} } \left( d\frac{ e^{-( d-1) t } + e^{-t} }{ ( 1- e^{-t} )^{d} 
  }-\frac{1+e^{-(d-2)t}}{(1-e^{-t})^{d-2}} \right). 
  \end{eqnarray}
  We can compare this expression for the co-exact $1$-form to the 
  partition function of the transverse traceless for $s=1$.  On comparing with (F.12) of 
   \cite{Anninos:2020hfj}  we see that we are missing the non-local term $\log Z_{\rm res}$ 
   which in the end 
   cancels off in (F.16). 
   We have obtained directly equation 
  (F.16) of \cite{Anninos:2020hfj} which is  local. 
  This reason for this may be attributed to 
  our different starting point in (\ref{detegi2}). 
  \cite{Anninos:2020hfj} did not have the second term which resulted from the 
  use of the representation of the logarithm in (\ref{logiden}). 
  As a consistency check of our approach we repeat the analysis 
  for rank $s$ symmetric traceless tensors in appendix \ref{spin-s-sum} and obtain the same conclusions. 
  It is interesting that our starting point directly gives the representation of the 
  path integral for the co-exact  $p$-forms as well as transverse traceless tensors directly without any 
  non-local terms. It would also be interesting to repeat the analysis using the 
  starting point of  \cite{Anninos:2020hfj} for the co-exact $p$-forms and reproduce our results, it 
  would require the identification of the number of killing tensors of these forms. 
  
  Let us  now examine the  entire  bulk contribution to the partition function of the $1$-form using the expression for the 
  path integral in (\ref{gfpform}) which includes the ghosts.
  We obtain 
  \begin{eqnarray}\label{naivchr}
  \left. \log {\mathcal Z}_1[S^{d+1} ]\right|_{\rm bulk}  &=&
    \int_{0}^{\infty}\frac{dt}{2t}\frac{ 1+ e^{-t}}{ 1- e^{-t} } \left( d\frac{ e^{-( d-1) t } +e^{-t} }{ ( 1- e^{-t} )^{d} }
    - \frac{ e^{-d t} + 1 }{ ( 1- e^{-t} )^d}  \right)
  \end{eqnarray}
  Thus we see  that the character in the integrand can be identified 
   with the  $\hat \chi_{{\rm bulk}, \, s}$ in equation 
  (G.18) for \cite{Anninos:2020hfj}, which is the naive bulk character. 
  Now to obtain the `flipped' character \footnote{We will review the procedure of obtaining the 
  `flipped' character in the examples discussed subsequently.}
   from  (\ref{naivchr}) we need to subtract the coefficient which contributes at the $t\rightarrow \infty$ limit from the term in the curved bracket.  This results  in 
  \begin{eqnarray} \label{naivchr2}
  \left. \log \widetilde{\mathcal Z}_1[S^{d+1} ]\right|_{\rm bulk}  &=&
    \int_{0}^{\infty}\frac{dt}{2t}\frac{ 1+ e^{-t}}{ 1- e^{-t} } \left( d\frac{ e^{-( d-1) t } +e^{-t} }{ ( 1- e^{-t} )^{d} }
    - \frac{ e^{-d t} + 1 }{ ( 1- e^{-t} )^d}   +1 \right)
  \end{eqnarray}
  The procedure of `flipping' does not change the coefficient of the $1/t$ term for even $d+1$ or 
  the IR finite term for odd $d+1$ as discussed  subsequently.  However 
  as noted  in \cite{Anninos:2020hfj} the expression in the curved bracket of 
  (\ref{naivchr2}) coincides with the character of the  unitary irreducible representation  of 
  $SO(1, d+1)$ 
  belonging to the exceptional series which is called the Harish-Chandra character of the 
  massless $1$-form or the rank one tensor.  It is an interesting question whether the 
  `flipped'  bulk character coincides with for all $p$-forms coincides with characters of  unitary irreducible 
  representation of \cite{Anninos:2020hfj} belonging to the exceptional series. 
  As a small step we perform this check for the $2$-form in $d+1=6$ below.

Finally note that 
the character sum in the integrand  of (\ref{chrintrep}) 
runs all the way over to the $0$-form in 
$d-2p$ dimensions. Therefore if $2p>d$, we would have terms which grow in $t$ exponentially. 
We should then  think of these characters as naive characters and convert them to flipped characters
using the rules given in \cite{Anninos:2020hfj}.

\begin{figure}[h]
\centering
\begin{tikzpicture}[thick,scale=0.85]
\filldraw[magenta] 
                (0,0.5) circle[radius=3pt]
                (0,-0.5) circle[radius=3pt];
                \filldraw[green] 
                (0.1,0) circle[radius=2pt]
                (0.5,0) circle[radius=2pt]
              (1,0) circle[radius=2pt]  
              (1.5,0) circle[radius=2pt]
              (2,0) circle[radius=2pt]
               (2.5,0) circle[radius=2pt]
                (3,0) circle[radius=2pt]
               (3.5,0) circle[radius=2pt]
                (4,0) circle[radius=2pt]
                (-0.1,0) circle[radius=2pt]
                (-0.5,0) circle[radius=2pt]
              (-1,0) circle[radius=2pt]  
              (-1.5,0) circle[radius=2pt]
              (-2,0) circle[radius=2pt]
               (-2.5,0) circle[radius=2pt]
                (-3,0) circle[radius=2pt]
               (-3.5,0) circle[radius=2pt]
                (-4,0) circle[radius=2pt] ;
\draw [decorate,decoration=snake] (0,-0.5) -- (0,-4);
\draw [decorate,decoration=snake] (0,0.5) -- (0,4);
\draw [postaction={decorate,decoration={markings , 
mark=at position 0.55 with {\arrow[black,line width=0.5mm]{<};}}}](0.2,1) arc[start angle=0, end angle=-180, radius=0.2cm];
\draw
[
postaction={decorate,decoration={markings , 
mark=at position 0.20 with {\arrow[red,line width=0.5mm]{>};}}}
][red, thick] (-4,0.2)--(0,0.2);
\draw[red, thick] (0,0.2)--(4,0.2);
\draw (0.4,1) node{$\mathbf{\epsilon}$};
\draw (0.4,-1) node{$\mathbf{-\epsilon}$};
\draw[gray, thick] (0,0) -- (0,4);
\draw[gray, thick] (0,0) -- (0,-4);
\draw[gray, thick] (0.2,4) -- (0.2,1);
\draw[gray, thick] (-0.2,1) -- (-0.2,4);
\draw[gray, thick] (0,0) -- (4,0);
\draw[gray, thick] (0,0) -- (-4,0);
\draw[gray, thick] (2,3) -- (2,3.4);
\draw[gray, thick] (2,3) -- (2.5,3);
\draw (2.3,3.3) node{$\mathbf{u}$};
\draw [red,thick](-2,0.5) node{$\mathbf{C}$};
\draw [gray,thick](0.5,3) node{$\mathbf{C'}$};
\end{tikzpicture}
\caption{Integration contour in $u$-plane} \label{fig1}
\qquad
\centering
\begin{tikzpicture}[thick,scale=0.85]
\filldraw[magenta] 
                (0.5,0) circle[radius=3pt]
                (-0.5,0) circle[radius=3pt];
                \filldraw[green] 
               ( 0,0.2) circle[radius=2pt]
                (0,0.5) circle[radius=2pt]
              (0,1) circle[radius=2pt]  
              (0,1.5) circle[radius=2pt] 
              (0,2) circle[radius=2pt] 
              (0,2.5) circle[radius=2pt]
                (0,3) circle[radius=2pt]
               (0,3.5) circle[radius=2pt]
                (0,4) circle[radius=2pt]
                (0,-0.1) circle[radius=2pt]
                (0,-0.5) circle[radius=2pt]
              (0,-1) circle[radius=2pt]  
              (0,-1.5) circle[radius=2pt]
              (0,-2) circle[radius=2pt]
               (0,-2.5) circle[radius=2pt]
                (0,-3) circle[radius=2pt]
               (0,-3.5) circle[radius=2pt]
                (0,-4) circle[radius=2pt] ;
\draw [decorate,decoration=snake] (-0.5,0) -- (-4,0);
\draw [decorate,decoration=snake] (0.5,0) -- (4,0);
\draw
[
postaction={decorate,decoration={markings , 
mark=at position 0.20 with {\arrow[red,line width=0.5mm]{>};}}}
][blue, thick] (0.5,0.2)--(4,0.2);
%\draw (7.9,0) node {$\Re(w)$};
%\draw (0,6.5) node {$\Im(w)$};
%\draw (5.5,2.5) node {$\Re(w)>0$};
%\draw (-5.5,2.5) node{$\Re(w)<0$};
%\draw (3,-3) node{$|w|=1$};
\draw (1,0.4) node{$\mathbf{\epsilon}$};
\draw (-1,0.4) node{$\mathbf{-\epsilon}$};
\draw[gray, thick] (0,0) -- (0,4);
\draw[gray, thick] (0,0) -- (0,-4);
\draw[gray, thick] (0,0) -- (4,0);
\draw[gray, thick] (-4,0) -- (0,0);
\draw[gray, thick] (2,3) -- (2,3.4);
\draw[gray, thick] (2,3) -- (2.5,3);
\draw (2.3,3.3) node{$\mathbf{t}$};
\end{tikzpicture}
\caption{Integration contour in $t$-plane} \label{fig2}
\end{figure}

Let us now apply the expression (\ref{chrintrep}) to obtain character integral representations 
of  the one loop determinants of the $p$-form. 

\subsection*{1-form on $S^4$}

From the expression  given in  (\ref{gfpform})
for the gauge fixed determinant of the 1-form on $S^4$, the one loop partition function  is given by 
\begin{eqnarray}
\log{\cal Z}_{1} [S^4] = -\frac{1}{2}  \log ( {\rm det}_T \Delta_1^{S^{4} })   + 
\frac{1}{2} \log ( {\rm det}' \Delta_0^{S^4} ) .
\end{eqnarray}
Here we have ignored the dependence  on the radius of the sphere. 
Substituting the expression in  (\ref{chrintrep}) we  obtain
\begin{eqnarray}
\log{\cal Z}_{1} [S^4] 
    &=&\int_{0}^{\infty}\frac{dt}{2t}\frac{1+e^{-t}}{1-e^{-t}}\left(\sum_{i=0}^{1}(-1)^i\chi^{dS}_{(3-2i,1-i)} (t) 
    -\chi^{dS}_{(3,0) (t) }\right), \\ \nonumber
    &=& -\int_{0}^{\infty}\frac{dt}{2t}\left(\frac{2 \left(e^{-t}+1\right)^2 \left(-3 e^{-t}+e^{-2 t}+1\right)}{\left(1-e^{-t}\right)^4}\right).
\end{eqnarray}
The second line in the above equation is obtained by substituting the explicit values of the characters 
given in (\ref{defchr}). 
The coefficient of the logarithmic divergence which is a renormalization group invariant 
can be obtained by examining the coefficient of $\frac{1}{t}$ of the integrand. 
This is given by 
\begin{eqnarray}
\log{\cal Z}_{1} [S^4] |_{\rm log\;divergence} = -\frac{31}{45}.
\end{eqnarray}
As expected this agrees with the trace anomaly coefficient of the 1-form.
\footnote{ There are several references from which this coefficient can be read out from. 
One reference is  \cite{Raj:2016zjp}.  This paper quotes values for $\tilde F = 
 (-1)^{\frac{d+1}{2} } F $ for even $d+1$.  Note that we are looking at the $\log Z =-F$. }

\subsubsection*{2-form on $S^6$}

From  (\ref{gfpform}), we see that partition function of the 2-form is given by the following combination 
of the determinant of co-exact forms
\begin{eqnarray} \label{2fs6}
\log{\cal Z}_{2} [S^6] = -\frac{1}{2}  \log ( {\rm det}_T \Delta_2^{S^{6} })   + 
\frac{1}{2}  \log ( {\rm det}_T \Delta_1^{S^{6} }) -
\frac{1}{2} \log ( {\rm det}' \Delta_0^{S^6} ). 
\end{eqnarray}
Again we have ignored the dependence on the radius. 
Using the character integral representation of the partition function of the co-exact form
we get
\begin{equation}
  \log{\cal Z}_{2} [S^6]   = 
  \int_{0}^{\infty}\frac{dt}{2t}\frac{1+e^{-t}}{1-e^{-t}}(\sum_{i=0}^{2}(-1)^i\chi^{dS}_{(5-2i,2-i)}(t) -\sum_{i=0}^{1}(-1)^i\chi^{dS}_{(5-2i,1-i)}(t)+\chi^{dS}_{(5,0)}(t) ).
\end{equation}
Finally substituting the  Harish-Chandra characters  from (\ref{chrintrep}) we obtain
\begin{equation}
 \log{\cal Z}_{2} [S^6]  =
 \int_{0}^{\infty}\frac{dt}{2t}\frac{\left(e^{-t}+1\right)^2 \left(3 e^{-4 t}-16 e^{-3 t}+32 e^{-2 t}-16 e^{-t}+3\right)}{\left(1-e^{-t}\right)^6}.
\end{equation}
The trace anomaly of the conformal 2-form can be read out easily from the coefficient of the $\frac{1}{t}$ 
term which is given by 
\begin{equation}
\left.  \log{\cal Z}_{2} [S^6] \right|_{{\rm log \; divergence} } =   \frac{221}{210}.
 \end{equation}
 This again agrees with coefficient of the trace anomaly for the conformal 2-form in literature, 
 see \cite{Raj:2016zjp} 

Let us verify that if indeed the bulk character for the $2$-form coincides with that of 
the character of  the corresponding 
unitary irreducible representation (UIR)  of $SO(1,  6)$  in the exceptional series. 
According to the notation of \cite{Basile:2016aen}, the $2$-form coincides with $p= 3, \Delta = 3, r=2, \vec x =0$ 
in the exceptional series \footnote{This `$p$' refers to 
the $p$ used in \cite{Basile:2016aen} not the $p$-form. 
In general for a given $p$-form, the `$p$' of \cite{Basile:2016aen} is
$p+1$. }. It corresponds to a Young tableaux of a single column with 
$2$ boxes or $Y_3 = ( 1, 1)$.  The contribution to the character comes from the second 
summation of equation (13) of \cite{Basile:2016aen} .  
We need to correct this by a factor of $2$ as pointed out 
in \cite{Anninos:2020hfj}. 
With this correction, the character of the UIR   is given by 
\begin{equation}\label{chruir}
\chi^{dS} ( q = e^{-t} ,  \vec x = 0)  = 2 \frac{  q^5 - 5 q^4 + 10 q^3}{ ( 1- q)^5} 
\end{equation}
Here the coefficients in the numerator are the dimensions corresponding to the  singlet, 
a vector and an anti-symmetric tensor of $SO(5)$.  The overall factor of $2$ is the correction 
noted by \cite{Anninos:2020hfj} which agrees with the older results of \cite{10.3792/pja/1195522333}. 
Now let us examine the bulk character  of the $2$-form in from our analysis, 
we  use (\ref{2fs6}) and extract the bulk contribution from each of the determinants.
\begin{eqnarray}
\left. \log{\cal Z}_{2} [S^6] \right|_{\rm bulk} =  \int_{0}^\infty \frac{dt}{2t} 
\frac{1+e^{-t}}{1-e^{-t}} \left[ \binom{5}{2} \frac{ e^{-2t} + e^{-3t}}{( 1- e^{-t} )^5} 
- \binom{5}{1} \frac{ e^{-t} + e^{-4t}}{( 1- e^{-t} )^5} 
+ \binom{5}{0}  \frac{ 1 + e^{-5t}}{( 1- e^{-t} )^5} 
\right] \nonumber \\
\end{eqnarray}
We now 
subtract the coefficient which contributes at the $t\rightarrow \infty$ limit from the term in the square bracket to go over to the flipped character. We obtain 
\begin{eqnarray}
\left. \log\widetilde{\cal Z}_{2} [S^6] \right|_{\rm bulk} =  \int_{0}^\infty \frac{dt}{2t} 
\frac{1+e^{-t}}{1-e^{-t}} \left[ \binom{5}{2} \frac{ e^{-2t} + e^{-3t}}{( 1- e^{-t} )^5} 
- \binom{5}{1} \frac{ e^{-t} + e^{-4t}}{( 1- e^{-t} )^5} 
+ \binom{5}{0}  \frac{ 1 + e^{-5t}}{( 1- e^{-t} )^5}  -1
\right] \nonumber \\
\end{eqnarray}
Simplifying the terms in the square bracket we obtain
\begin{eqnarray} \label{flipbulk}
\left. \log\widetilde{\cal Z}_{2} [S^6] \right|_{\rm bulk} =  \int_{0}^\infty \frac{dt}{2t} 
\frac{1+e^{-t}}{1-e^{-t}} \left(  \frac{ 20 e^{-3t} - 10 e^{-4t} + 2 e^{-5t} }{ ( 1- e^{-t})^5} 
\right) 
\end{eqnarray}
We see that terms in the curved bracket of (\ref{flipbulk}) precisely coincides with  the 
character of the UIR  (\ref{chruir}) corresponding to the $2$-form in $d+1= 6$. 
It will be interesting to verify whether such a statement holds
 for all $p$-forms in arbitrary dimensions,  as it was demonstrated for the 
 case of massless symmetric tensors  in \cite{Anninos:2020hfj}
 \footnote{We  have also verified the fact that the UIR of the
 $2$-form in $d+1 = 7$ coincides with the corresponding flipped character.}.

\subsubsection*{2-form on $S^4$}

Let us now consider the case of $2$-form on $S^4$ for which $2p > d$.  From 
(\ref{gfpform}) we see that the partition function is given by 
\begin{eqnarray}
\log{\cal Z}_{2} [S^4] = -\frac{1}{2}  \log ( {\rm det}_T \Delta_2^{S^{4} })   + 
\frac{1}{2}  \log ( {\rm det}_T \Delta_1^{S^{4} }) -
\frac{1}{2} \log ( {\rm det}' \Delta_0^{S^4} ) .
\end{eqnarray}
Using the character representation  in (\ref{chrintrep}) for each of the co-exact $p$-forms that occur 
in the above expression we obtain 
\begin{eqnarray}\label{2fs4}
\log{\cal Z}_{2} [S^4]  = 
\int_{0}^{\infty}\frac{dt}{2t}\frac{1+e^{-t}}{1-e^{-t}}\left(\sum_{i=0}^{2}(-1)^i\chi^{dS}_{(3-2i,2-i)}(t)-\sum_{i=0}^{1}(-1)^i\chi^{dS}_{(3-2i,1-i)}(t)+\chi^{dS}_{(3,0)} (t) \right) .\nonumber \\
\end{eqnarray}
Note that the characters that occur for the co-exact 2-form on $S^4$  are given by 
\begin{equation}
  \chi^{dS}_{(3-2i,2-i)}= \binom{3-2i}{2-i}\frac{ \left(e^{-(2-i) t}+e^{-(1-i)t}\right)}{\left(1-e^{-t}\right)^{3-2i}},
\end{equation}
with $i$ running from $0$ to $2$ at $i = 2$, this naive character which grow as $e^{t}$ and therefore 
cannot be considered as  character of a unitary representation of $SO(1, d+1)$. 
This feature also occurred in \cite{Anninos:2020hfj} for character representation of one loop determinants 
of massless higher spin fields with  spins $\geq 2$. 
We can follow the same procedure developed in \cite{Anninos:2020hfj} to deal with such naive characters. 
We replace the naive character by the flipped character. 
Let $x = e^{-t}$, and consider the character $\chi = \sum_k c_k x^k  $ with terms $k<0$, then the 
flipped character is  given by 
\begin{equation} \label{flip}
[\chi]_+ =  \chi - c_0 - c_k ( x^ k + x^{ -k} ) .
\end{equation}
As explained in \cite{Anninos:2020hfj}, this procedure  can  be thought of as a contour deformation 
so that the integration is done over the negative $t$ axis and removing zero modes from the path integral. 

Let us demonstrate this for the character corresponding to the  co-exact 2-form.
\begin{eqnarray}
   \chi  &=&\sum_{i=0}^{2}(-1)^i\chi^{dS}_{(3-2i,2-i)}(t) \\ \nonumber
&=& \frac{\binom{3}{2} \left(x^2+x\right)}{(1-x)^3}-\frac{\binom{1}{1} (x+1)}{1-x}+\frac{\binom{-1}{0} \left( x^{-1}+1\right)}{(1-x)^{-1} }, \\ \nonumber
&=& \frac{1}{x}  -1  +10 x^2 + 25 x^3  + 46 x^4  + \cdots 
\end{eqnarray}
The flipped character is then given by
\begin{eqnarray}
[\chi]_+ &=& \chi - ( - 1) - \left(  x^{-1} + x \right) , \\ \nonumber
&=& \frac{ x ( -1 + 13 x - 8 x^2 +2 x^3) }{( 1- x)^3 }.
\end{eqnarray}

Note the integral with the flipped character has the same 
UV divergence as the original naive character.  The reason is that  the additional contributions 
to convert the naive character to the flipped character in (\ref{flip}) together with  
the factor $( 1+ x)/( 1-x)$ in the integral transform is an odd function of $t$ and therefore will 
not contribute to the coefficient of $\frac{1}{t}$.  

Extracting the logarithmic divergence from (\ref{2fs4})  we obtain
\begin{equation}
\left. \log{\cal Z}_{2} [S^4]\right|_{\rm log \; divergence } =  \frac{209}{90}.
\end{equation}
This coefficient agrees with the result quoted in \cite{Raj:2016zjp}.

\subsection{Logarithmic  divergence  on even spheres} \label{sectionlogdiv}

Proceeding as described in the previous section we can extract the coefficient of the 
logarithmic divergence for $p$ forms on spheres in even $d+1$ with $ 2\leq ( d+1) \leq 14$. 
The result is summarised in table \ref{tab1}

\begin{table}[ht]\renewcommand{\arraystretch}{2}
 \begin{tabular*}{.953\textwidth}{| l| l | l | l | l | l| l|l|}
 \hline
\diagbox[width=4em]{$p$}{$d+1$} &$2$& $4$ & $6$ & $8$ & $10$& $12$ & $14$ \\
  \hline\hline
 $0 $ &$\frac{1}{3}$ & $\frac{29}{90}$ & $\frac{1139}{3780}$ & $\frac{32377}{113400}$ & $\frac{2046263}{7484400}$& $\frac{5389909963}{20432412000}$ & $\frac{31374554287}{122594472000}$\\
  \hline\hline
 $1 $ &$ -2$ & $ -\frac{31}{45}$ & $ -\frac{1271}{1890}$ & $ -\frac{4021}{6300}$ & $- \frac{456569}{748440}$ & $- \frac{1199869961}{2043241200}$ & $ -\frac{893517041}{1571724000}$\\
  \hline\hline
 $2 $&  $ +4 $ & $ +\frac{209}{90}$ & $ +\frac{221}{210}$ & $ +\frac{2603}{2520}$ & $+\frac{13228}{13365}$ & $+\frac{1296877349}{1362160800}$ & $ +\frac{2686807471}{2918916000}$\\
  \hline\hline
 $3$& $  $ & $ -4$ & $-\frac{5051}{1890}$ & $ -\frac{8051}{5670}$ & $-\frac{5233531}{3742200}$ & $-\frac{1417811}{1051050}$ & $ -\frac{456732097}{350269920}$\\
  \hline\hline
$4$&  $  $ & $+6 $ & $+\frac{16259}{3780}$ & $ +\frac{7643}{2520}$ & $+\frac{1339661}{748440}$ & $+\frac{15630799}{8845200}$ & $ +\frac{933250433}{544864320}$\\
  \hline
  \hline
$5$& $  $ & $ $ & $ -6$ & $-\frac{29221}{6300}$ & $ -\frac{12717931}{3742200}$ & $-\frac{525793111}{243243000}$ & $-\frac{7280049421}{3405402000}$ \\
  \hline\hline
$6$&  $  $ & $ $ & $+8 $ & $+\frac{712777}{113400}$ & $ +\frac{66688}{13365}$ & $+\frac{33321199}{8845200}$ & $+\frac{3698905481}{1459458000}$ \\
  \hline\hline
 $7$& $  $ & $ $ & $ $ & $-8 $ & $-\frac{4947209}{748440}$ & $ -\frac{5622011}{1051050}$ & $-\frac{14090853421}{3405402000}$  \\
  \hline\hline
 $8$& $  $ & $ $ & $ $ & $ +10$ & $+\frac{61921463}{7484400}$ & $ +\frac{9469842149}{1362160800}$ & $+\frac{3112707713}{544864320}$  \\
  \hline\hline
 $9$& $  $ & $ $ & $ $ & $ $ & $-10 $ & $ -\frac{17545799561}{2043241200}$ & $-\frac{2558351617}{350269920}$  \\
  \hline\hline
 $10$& $  $ & $ $ & $ $ & $ $ &  $ +12$ & $ +\frac{209714029963}{20432412000}$& $+\frac{26038135471}{2918916000}$  \\
  \hline\hline
$11$&  $  $ & $ $ & $ $ & $ $ &  $ $ & $- 12 $ & $ -\frac{16610757041}{1571724000}$  \\
  \hline\hline
$12$&  $  $ & $ $ & $ $ & $ $ & $ $ & $+14 $ & $+\frac{1502508218287}{122594472000}$  \\
  \hline\hline
  $13$& $  $ & $ $ & $ $ & $ $ & $ $ & $ $ & $ -14$  \\
  \hline
\end{tabular*}
\caption{Logarithmic divergence in the partition function of $p$-forms on even spheres.}
\label{tab1}
\end{table}

Note that the $p$-forms which are Hodge dual to each other do not 
have the same logarithmic divergence in the partition function. 
The difference in this coefficient 
between such Hodge dual pair of $p$ forms have been seen to be integer multiples of $2$. 
Consider the $(0, 2)$-form pair on $S^4$, from table (\ref{tab1}) we see that the difference in the 
logarithmic divergence  is $-2$. Similarly the   coefficients for $(1, 3)$ and $(2, 4)$-form pairs on 
$S^6$ and $S^8$differ by $+2$  and $-2$ respectively. 
Now consider the $(0, 4)$, $(1, 4)$-pair on $S^6$ and  $S^8$, the coefficient of the 
logarithmic divergence differ by 
$-4$ and $4$ respectively. 
These jumps in the trace anomaly coefficients agree with that noted earlier in \cite{Raj:2016zjp}.

\subsection{IR finite term or the F-term on odd spheres} \label{sectionfterm}

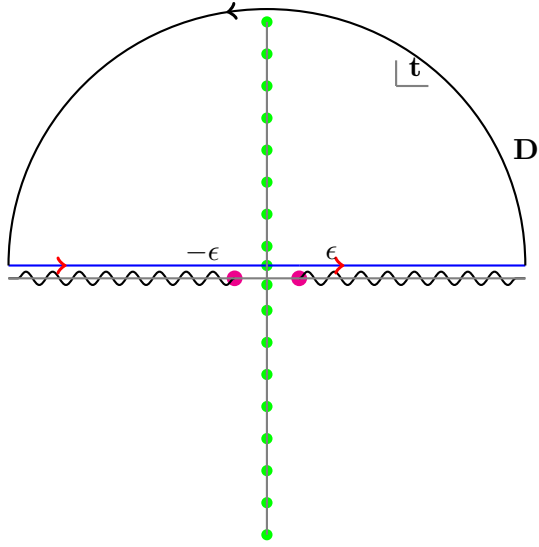
\begin{figure}
\centering
\begin{tikzpicture}[thick,scale=0.85]
\filldraw[magenta] 
                (0.5,0) circle[radius=3pt]
                (-0.5,0) circle[radius=3pt];
                \filldraw[green] 
               ( 0,0.2) circle[radius=2pt]
                (0,0.5) circle[radius=2pt]
              (0,1) circle[radius=2pt]  
              (0,1.5) circle[radius=2pt] 
              (0,2) circle[radius=2pt] 
              (0,2.5) circle[radius=2pt]
                (0,3) circle[radius=2pt]
               (0,3.5) circle[radius=2pt]
                (0,4) circle[radius=2pt]
                (0,-0.1) circle[radius=2pt]
                (0,-0.5) circle[radius=2pt]
              (0,-1) circle[radius=2pt]  
              (0,-1.5) circle[radius=2pt]
              (0,-2) circle[radius=2pt]
               (0,-2.5) circle[radius=2pt]
                (0,-3) circle[radius=2pt]
               (0,-3.5) circle[radius=2pt]
                (0,-4) circle[radius=2pt] ;
\draw [decorate,decoration=snake] (-0.5,0) -- (-4,0);
\draw [decorate,decoration=snake] (0.5,0) -- (4,0);
\draw
[
postaction={decorate,decoration={markings , 
mark=at position 0.20 with {\arrow[red,line width=0.5mm]{>};}}}
][blue, thick] (0.5,0.2)--(4,0.2);
\draw
[
postaction={decorate,decoration={markings , 
mark=at position 0.20 with {\arrow[red,line width=0.5mm]{>};}}}
][blue, thick] (-4,0.2)--(0.5,0.2);
%\draw (7.9,0) node {$\Re(w)$};
%\draw (0,6.5) node {$\Im(w)$};
%\draw (5.5,2.5) node {$\Re(w)>0$};
%\draw (-5.5,2.5) node{$\Re(w)<0$};
%\draw (3,-3) node{$|w|=1$};
\draw (4.01,2.01) node{$\mathbf{D}$};
\draw (1,0.4) node{$\mathbf{\epsilon}$};
\draw (-1,0.4) node{$\mathbf{-\epsilon}$};
\draw[gray, thick] (0,0) -- (0,4);
\draw[gray, thick] (0,0) -- (0,-4);
\draw[gray, thick] (0,0) -- (4,0);
\draw[gray, thick] (-4,0) -- (0,0);
\draw[gray, thick] (2,3) -- (2,3.4);
\draw[gray, thick] (2,3) -- (2.5,3);
\draw (2.3,3.3) node{$\mathbf{t}$};
\draw[postaction={decorate,decoration={markings , 
mark=at position 0.55 with {\arrow[black,line width=0.5mm]{>};}}}](4,0.2) arc[start angle=0, end angle=180, radius=4cm];
\end{tikzpicture}
\caption{Contour  for extracting  the contribution of IR finite term in odd spheres} \label{fig3}
\end{figure}

Consider the case $d+1$ is odd, the character integral representation of the 
determinant of the co-exact $p$ form is given by (\ref{chrintrep}) with the characters given in 
(\ref{defchr}).  Using the definition of the characters it is easy to see that for $d$ even, 
The characters are all even functions of $t$ which makes 
the integrand in (\ref{chrintrep}) is an even function since the remaining factor $( 1 + e^{-t}) / t( 1 - e^{-t}) $
is also an even function of $t$.
Therefore the contour in $t$ can be extended to the whole real line as shown in 
figure \ref{fig3}.  
Note that to ensure that the integrand is  IR finite one always replaces the character 
by the flipped character.  The additional terms  one adds to flip the character is always even and 
 the flipped character by construction converges as $t\rightarrow\infty$ in the 
complex plane. 
This enables us to close the contour using a large semi-circle  in either the upper half or lower half plane. 
In figure  \ref{fig3} we have chosen to close the contour $D$ in  the upper half plane. 
The integral then can evaluated by summing over the residues that occur on the imaginary axis. 
This results in the IR finite term. The IR finite term of the partition function is negative of what is 
known as the F-term in the literature.
We illustrate this  method in two examples.

\subsubsection*{ 2-form on $S^7$}

The character for the co-exact 2-form in $d=6$ is given by 
\begin{eqnarray}
\chi(t)  &=&  \sum_{i = 0}^2  (-1)^i  \chi^{dS}_{( 6 - 2i , 2-  i )  } ( t) , \\ \nonumber
&=& 1- 2x  + 3 x^2  + 52 x^3 + 242 x^4 + \cdots, \qquad x = e^{-t} 
\end{eqnarray}
Therefore the flipped character is given by 
\begin{eqnarray}
[\chi ( t) ]_+= \chi(t) -1.
\end{eqnarray}
The IR finite term is obtained by evaluating the integral
\begin{eqnarray}
\left. - \frac{1}{2}  \log ( {\rm det}_T \Delta_2^{S^{7} })  \right|_{{\rm IR \; finite}}
= \int_{D} \frac{dt}{4t} \frac{ 1+ e^{-t}}{ 1- e^{-t}} [\chi(t)]_+
\end{eqnarray}
Here $D$ is the contour in the upper half plane as shown in figure (\ref{fig3}) 
Then evaluating  the residues  we obtain 
\begin{eqnarray}
\left. - \frac{1}{2}  \log ( {\rm det}_T \Delta_2^{S^{7} })  \right|_{{\rm IR \; finite}}
%&=& \sum_{n=1}^\infty \frac[ - \frac{15}{ 64\pi^6 n^7} + \frac{1}{16\pi^4 n^5 } + \frac{9}{ 32 \pi^2 n^3} , 
%\\ \nonumber
&=&  \frac{9}{32 \pi^2} \zeta(3) + \frac{1}{16 \pi^4} \zeta( 5) - \frac{15}{64\pi^6} \zeta ( 7) .
\end{eqnarray}

\subsubsection*{ 2-form on $S^3$} 

The naive character for the co-exact $2$ form in $d=2$ is given by 
\begin{eqnarray}
\chi( t) &=&    \sum_{i = 0}^2  (-1)^i \chi^{dS}_{( 2 - 2i , 2- i )  } ( t) , \\ \nonumber
&=& \frac{1}{x^2} - \frac{2}{x}  + 3 + 5x^2  + 6 x^3  + 8 x^4 + \cdots 
\end{eqnarray}
Therefore the flipped character is given by 
\begin{eqnarray}
[\chi (t) ]_+ = \chi( t) - 3  + 2 ( e^{t} + e^{ - t} ) - ( e^{-2t} + e^{ -2t} ) .
\end{eqnarray}
Proceeding as before,  the IR finite contribution to the   one loop determinant is given by 
\begin{eqnarray}
\left. - \frac{1}{2}  \log ( {\rm det}_T \Delta_2^{S^{7} })  \right|_{{\rm IR \; finite}}
&=& \int_{D} \frac{dt}{4t} \frac{ 1+ e^{-t}}{ 1- e^{-t}} [\chi(t)]_+, \\ \nonumber
%=-  \sum_{n =1}^\infty \frac{1}{ 4 \pi^2 n^3} , \\ \nonumber
&=& - \frac{1}{ 4\pi^2} \zeta(3) .
\end{eqnarray}

We  use  the method described to obtain the values of of the IR finite terms for $p$ forms 
for $0\leq p\leq 7$   in $ 3\leq d+1 \leq 9$. 
We have compared the results of these finite terms wherever possible with the results 
of \cite{Raj:2016zjp} and noted that they agree. For instance the  results of the IR finite part evaluated
 in equations (2.29)-(2.34)  of \cite{Raj:2016zjp} agree with the corresponding values in 
table \ref{table2}. There are differences in sign, due to the fact that what is quoted in \cite{Raj:2016zjp}
is a quantity called $\tilde F = (-1)^{d/2} $ times the sphere free energy in odd $d+1$, while we have 
evaluated $\log{\cal Z}$ which is negative of the free energy. 
It is easy to observe  the IR finite part respects Hodge duality as was seen earlier in 
\cite{Giombi:2015haa,Raj:2016zjp}. For example the values of the IR finite part for the Hodge-dual 
pair  of forms $(0, 1)$  in $d =2$ agree.  The same property holds for the pairs
$(0, 3), ( 1, 2)$  in $d=4$, other Hodge-dual pairs can be seen in table \ref{table2}. 

 \begin{table}[ht]
\centering { \footnotesize{
\begin{tabular}{c|l}
\hline
$d+1 $ &  ~~~~~~~~~~~~~~~~~~~~~~~ $p=0$ \\
\hline 
\\
$3$ & \quad$-\frac{\zeta (3)}{4 \pi ^2} $ 
   \\   
\\
$5$ & \quad$-\frac{23 \zeta (3)}{48 \pi ^2}+\frac{\zeta (5)}{16 \pi ^4}$  \\
 \\
$7$ & \quad$\-\frac{949 \zeta (3)}{1440 \pi ^2}+\frac{13 \zeta (5)}{48 \pi ^4}-\frac{\zeta (7)}{64 \pi ^6} $   \\
 \\
 $9$ & \quad$-\frac{16399 \zeta (3)}{20160 \pi ^2}+\frac{2087 \zeta (5)}{3840 \pi ^4}+\frac{\zeta (9)}{256 \pi ^8}-\frac{31 \zeta (7)}{256 \pi ^6} $   \\
 \\
\hline
\end{tabular}
\vspace{5mm}
\quad
\begin{tabular}{c|l}
\hline
$d+1 $ &  ~~~~~~~~~~~~~~~~~~~~~~ $p=1$ \\
\hline 
\\
$3$ & \quad$-\frac{\zeta (3)}{4 \pi ^2} $ 
   \\   
\\
$5$ & \quad$\frac{5 \zeta (3)}{16 \pi ^2}+\frac{3 \zeta (5)}{16 \pi ^4}$  \\
 \\
$7$ & \quad$\frac{179 \zeta (3)}{288 \pi ^2}+\frac{17 \zeta (5)}{48 \pi ^4}-\frac{5 \zeta (7)}{64 \pi ^6} $   \\
 \\
 $9$ & \quad$\frac{841 \zeta (3)}{960 \pi ^2}+\frac{1249 \zeta (5)}{3840 \pi ^4}+\frac{7 \zeta (9)}{256 \pi ^8}-\frac{101 \zeta (7)}{256 \pi ^6} $   \\
 \\
\hline
\end{tabular}
\quad
\begin{tabular}{c|l}
\hline
$d+1 $ &  $~~~~~~~~~~~~~~~~~~~~p=2$ \\
\hline 
\\
$5$ & \quad$\frac{5 \zeta (3)}{16 \pi ^2}+\frac{3 \zeta (5)}{16 \pi ^4} $ 
   \\   
\\
$7$ & \quad$\-\frac{49 \zeta (3)}{144 \pi ^2}-\frac{7 \zeta (5)}{24 \pi ^4}-\frac{5 \zeta (7)}{32 \pi ^6}$  \\
 \\
$9$ & \quad$-\frac{1987 \zeta (3)}{2880 \pi ^2}+\frac{21 \zeta (9)}{256 \pi ^8}-\frac{2333 \zeta (5)}{3840 \pi ^4}-\frac{71 \zeta (7)}{256 \pi ^6} $   \\
 \\
 \\
\hline
\end{tabular}
\vspace{5mm}
\quad
\begin{tabular}{c|l}
\hline
$d+1 $ &   $~~~~~~~~~~~~~~~~~~~p=3$ \\
\hline 
\\
$5$ & \quad$\frac{\zeta (5)}{16 \pi ^4}-\frac{23 \zeta (3)}{48 \pi ^2} $ 
   \\   
\\
$7$ & \quad$-\frac{49 \zeta (3)}{144 \pi ^2}-\frac{7 \zeta (5)}{24 \pi ^4}-\frac{5 \zeta (7)}{32 \pi ^6}$  \\
 \\
$9$ & \quad$\frac{205 \zeta (3)}{576 \pi ^2}+\frac{91 \zeta (5)}{256 \pi ^4}+\frac{75 \zeta (7)}{256 \pi ^6}+\frac{35 \zeta (9)}{256 \pi ^8}$   \\
 \\
 \\
\hline
\end{tabular}
\quad
\begin{tabular}{c|l}
\hline
$d+1 $ &  $~~~~~~~~~~~~~~~~~~p=4$ \\
\hline 
\\
$7$ & \quad$\frac{179 \zeta (3)}{288 \pi ^2}+\frac{17 \zeta (5)}{48 \pi ^4}-\frac{5 \zeta (7)}{64 \pi ^6}$ 
   \\   
\\
$9$ & \quad$\frac{205 \zeta (3)}{576 \pi ^2}+\frac{91 \zeta (5)}{256 \pi ^4}+\frac{75 \zeta (7)}{256 \pi ^6}+\frac{35 \zeta (9)}{256 \pi ^8}$  \\
 \\ \\
\hline
\end{tabular}
\quad
\vspace{5mm}
\begin{tabular}{c|l}
\hline
$d+1 $ &  ~~~~~~~~~~~~~~~~~~~~~~~ $p=5$ \\
\hline 
\\
$7$ & \quad$-\frac{949 \zeta (3)}{1440 \pi ^2}+\frac{13 \zeta (5)}{48 \pi ^4}-\frac{\zeta (7)}{64 \pi ^6}$ 
   \\   
\\
$9$ & \quad$-\frac{1987 \zeta (3)}{2880 \pi ^2}+\frac{21 \zeta (9)}{256 \pi ^8}-\frac{2333 \zeta (5)}{3840 \pi ^4}-\frac{71 \zeta (7)}{256 \pi ^6}$  \\
 \\ \\
\hline
\end{tabular}
\quad
\begin{tabular}{c|l}
\hline
$d+1 $ & ~~~~~~~~~~~~~~~~~~~~~~ $p=6$ \\
\hline 
\\
$9$ & \quad$\frac{841 \zeta (3)}{960 \pi ^2}+\frac{1249 \zeta (5)}{3840 \pi ^4}+\frac{7 \zeta (9)}{256 \pi ^8}-\frac{101 \zeta (7)}{256 \pi ^6}$ 
   \\   \\
\hline
\end{tabular}
\quad
\begin{tabular}{c|l}
\hline
$d+1 $ & ~~~~~~~~~~~~~~~~~~~~~~~~~~~~ $p=7$ \\
\hline 
\\
$9$ & \quad$-\frac{16399 \zeta (3)}{20160 \pi ^2}+\frac{2087 \zeta (5)}{3840 \pi ^4}+\frac{\zeta (9)}{256 \pi ^8}-\frac{31 \zeta (7)}{256 \pi ^6}$ 
   \\   
\\
\hline
\end{tabular}
\caption{IR finite part of the partition function on odd spheres.}
\label{table2}
}}
\end{table}

\section{$p$-forms on anti-de Sitter space} \label{section2}

The construction of the  partition function of gauge fixed $p$-form field on $AdS_{d+1} $ 
proceeds along the same lines  as that done for the sphere. 
The partition function is given by 
\begin{align} 
    \mathcal{Z}_p[AdS_{d+1}] &=\Big[\frac{1}{\det_T \Delta_p}\frac{\det_T\Delta_{p-1}}{\det_T\Delta_{p-2}}\cdots\big(\frac{\det_T\Delta_1}{\det\Delta_0})^{(-1)^p}\Big]^{\frac{1}{2}} \label{p-fads},
\end{align}
where
$\det_T\Delta_p$ denotes the determinant of Hodge de-Rham Laplacian of co-exact $p$-form field
 and $\det\Delta_0$  is the determinant of $0$-form on $AdS$. 
 Note that the 0-form Laplacian does not have a discrete zero mode unlike the case 
 for spheres. Therefore the one loop path integral does not contain the volume of 
 $AdS$. 
 The expression implies again that the key 
  ingredient to evaluate the partition function is the determinant of the  co-exact $p$-forms on $AdS$.
  In section  \ref{detco2} we will show that  these determinants can be written 
  in terms of Harish-Chandra characters of  the anti-de Sitter group. 
  In section \ref{section2.2} we evaluate the trace anomaly coefficient in  $AdS_{d+1}$ with $d+1$ even and show 
  that it is proportional to that of $S^{d+1}$ as expected since these spaces are conformally flat. 
    
 \subsection{Determinant of co-exact $p$-forms as character integrals}
 \label{detco2}
 
  To evaluate determinants, we need both the eigen values and their degeneracies of the 
  Hodge-deRham Laplacian. 
  Since $AdS$ is non-compact, the eigen values are part of continuous spectrum  distributed through a 
  measure known as the Plancherel measure. 
  These eigen values  of the Laplacian are  given by \cite{CAMPORESI199457}
\begin{align}
    \Delta_p \psi_{\lambda}^{(u_i)}=-(\lambda^2+(\frac{d }{2}-p)^2)\psi_{\lambda}^{(u_i)},
\end{align}
where $\lambda$ runs from $0$ to $\infty$  and $\psi_\lambda^{(u_i)}$ is the basis of eigen functions for co-exact 
$p$-forms. 
The Plancherel  measure is given by \cite{CAMPORESI199457}
\begin{eqnarray} \label{defmup}
\mu_p(\lambda) = 
N(d)\hat{g}(p) \times 
\begin{cases}
\frac{\lambda  \tanh (\pi  \lambda ) \prod _{j=\frac{1}{2}}^{\frac{d}{2}} \left(j^2+\lambda ^2\right)}{\left(\frac{d}{2}-p\right)^2+\lambda ^2},  \qquad  & \hbox{for}\;  d  \; \hbox{odd}   \\
\frac{ \prod _{j=0}^{\frac{d}{2}} \left(j^2+\lambda ^2\right)}{\left(\frac{d}{2}-p\right)^2+\lambda ^2},  
\qquad  & \hbox{for} \; d\; \hbox{even}
\end{cases}
\end{eqnarray}
where the product runs over half integers for $d$ odd and integers for $d$ even. 
The normalizations are given by 
\begin{equation} \label{defnd}
N(d) =  \frac{ {\rm Vol}( AdS_{d+1} ) }{2^d \Gamma( \frac{d+1}{2} ) \pi^{\frac{d+1}{2} } }, 
\qquad 
\hat g(p ) = \frac{d!}{p! (d -p)!}.
\end{equation}
By ${\rm Vol } (AdS_{d+1})$ we refer to the regularised volume given by 
\begin{eqnarray}\label{regvol}
{\rm Vol } ( AdS_{d+1}) = 
\begin{cases}
\pi^{\frac{d}{2} } \Gamma(  - \frac{d}{2} ) ,  \qquad  & \hbox{for}\;  d  \; \hbox{odd}   \\
\frac{ 2 ( -\pi)^{\frac{d}{2}  }}{ \Gamma( \frac{d}{2} + 1) } \log \tilde R
\qquad  & \hbox{for} \; d\; \hbox{even}.
\end{cases}
\end{eqnarray}
Here $\tilde R$ is the dimensionless IR cutoff, the ratio fo the radial cutoff to the radius of $AdS$.

Using these inputs let us proceed to evaluate the determinant of the co-exact $p$-form 
\begin{eqnarray}
-\frac{1}{2}\log(  {\rm det}_T \Delta_p^{AdS_{d+1} } ) = - \frac{1}{2} \int_0^\infty
 d\lambda \mu_p(\lambda) \log \left[  \lambda^2  +(  \frac{d}{2}  -p)^2  \right].
\end{eqnarray}
We again replace the logarithm by the identity in (\ref{logiden}) to obtain
\begin{equation}
-\frac{1}{2}\log(  {\rm det}_T \Delta_p^{AdS_{d+1} } ) = 
\int_0^\infty \frac{d\tau}{ 2\tau}  \int_0^\infty d\lambda  \mu_p (\lambda) ( 
e^{-\tau ( \lambda^2 + ( \frac{d}{2} - p )^2 ) } - e^{-\tau} ) .
\end{equation}
The integral over the Plancherel  measure vanishes
\begin{equation}\label{sumpla}
\int_0^\infty d\lambda   \mu_p (\lambda) = 0.
\end{equation}
Just as in the case of equation (\ref{sumdegen}) for spheres, 
we obtain this result  in (\ref{sumpla}) by  choosing $d$ to be sufficiently negative and then 
analytically continuing the result to  positive $d$ \footnote{Since $\mu_p(\lambda)$ is an even function of 
$\lambda$, it is equivalent to show $\int_{-\infty}^\infty d\lambda \mu_p (\lambda ) = 0$. 
By replacing $\mu_p(\lambda) $ with its Fourier transform
$W_p(u)$ and performing the integration over $\lambda$ we see that we get 
$\int_{-\infty}^\infty d\lambda \mu_p (\lambda ) = \lim_{u\rightarrow 0 } W_p(u)$. 
From  (\ref{niceform}) we see that in this limit $W_p(u)$ vanishes for sufficiently large negative $d$. }.
Thus we can proceed by regulating only the first term which results in 
\begin{equation}\label{man1} 
-\frac{1}{2}\log(  {\rm det}_T \Delta_p^{AdS_{d+1} } ) 
= \int_0^\infty \frac{d\tau}{ 4\tau} e^{ - \frac{\epsilon^2}{4\tau} } 
\int_{-\infty} ^\infty d\lambda  \mu_p( \lambda)   e^{-\tau ( \lambda^2 + ( \frac{d}{2} - p )^2 )}.
\end{equation}
Here we have used the fact that the Plancherel measure is symmetric in $\lambda$. 
We now write the Plancherel measure in terms of its Fourier transform. 
\begin{eqnarray}\label{defft}
\mu_p(\lambda) = \int_C \frac{du}{2\pi} e^{ - i \lambda u } W_p( u ) .
\end{eqnarray}
Here $C$ is a contour chosen to ensure the Fourier transform is well defined, which will be detailed 
subsequently.  The contour differs when  $d+1$ is even or odd. We will denote this  as
$C_e, C_o$ respectively. 
Substituting (\ref{defft}) in ( \ref{man1}) and performing the integration over $\lambda$, 
we obtain 
\begin{equation}\label{man11}
-\frac{1}{2}\log(  {\rm det}_T \Delta_p^{AdS_{d+1} } )  = \int_C  du
\int_0^\infty \frac{d\tau}{ 8 ( \pi \tau^3)^{\frac{1}{2}}  }
e^{ - \frac{\epsilon^2 + u^2}{4\tau} - \tau   ( \frac{d}{2} - p )^2  } W_p( u ) .
\end{equation}
After integration over $\tau$ we obtain 
\begin{equation} \label{gdet}
-\frac{1}{2}\log(  {\rm det}_T \Delta_p^{AdS_{d+1} } )  = 
\int_C \frac{du}{ 4 \sqrt{ u^2 + \epsilon^2}} e^{  - ( \frac{d}{2} - p ) \sqrt{ u^2 + \epsilon^2}  }  W_p(u) .
\end{equation}
We can take the $\epsilon\rightarrow 0$ limit at the end \footnote{After performing the 
integration over $\tau$ one gets the factor $|\frac{d}{2} -p|$ in the exponent. We have chosen one branch since we expect the answer to be analytic in $d, p$.  This will be seen in the subsequent analysis. }.
Thus the task of determining the one loop determinant is reduced to finding the 
Fourier transform $W_p(u)$.

\subsubsection*{Fourier transform of the Plancherel measure}

To construct the  Fourier transform $W_p(u)$, 
consider the ratio of the Plancherel measure of the $p$-form to the $0$-form. 
From (\ref{defmup}) we see that this is given by 
\begin{align}\label{rel0p}
    \frac{\mu_p(\lambda)}{\mu_0(\lambda)}=\hat{g}(p)\frac{\lambda^2+(\frac{d}{2})^2}{\lambda^2+(\frac{d}{2}-p)^2}.
\end{align}
From the definition of the Fourier transform in (\ref{defft}), the inverse  is given by 
\begin{equation} \label{invft}
W_p(u) = \int_{-\infty}^\infty  d\lambda e^{i \lambda u } \mu_p(\lambda) ,
\end{equation}
where the integral is on the real line. 
Using the relation 
in  (\ref{rel0p}),  
we obtain the following 
the following differential equation satisfied by 
$W_p(u)$. 
\begin{equation}\label{diffrel}
     \frac{d^2}{d u^2}W_p(u)-(p-\frac{d}{2})^2W_p(u)=\hat{g}(p)\left(\frac{d^2W_0(u)}{d u^2}-\frac{d^2}{4} W_0(u)\right).
\end{equation}
Here we have assumed that the derivative with respect to $u$ can be taken inside the integral
involved in the defining the Fourier transform. 
$W_0(u)$ has been constructed in \cite{Sun:2020ame}, 
On substituting $W_0(u)$ the  differential equation  in (\ref{diffrel}) 
becomes an inhomogenous second order ordinary differential equation for $W_p(u)$. 

Before we solve the differential equation, let us recall the construction of $W_0(p)$ which was 
done in \cite{Sun:2020ame}. 
For both even and odd $d$ it is given by 
\begin{eqnarray} \label{wo}
W_0(u) &=&  \frac{ 1+ e^{-u}}{ 1- e^{-u}} \frac{ e^{-\frac{d}{2} u }}{ ( 1- e^{-u} )^d} .
\end{eqnarray}
However the contour relating $W_0(u)$ to the Plancherel measure in (\ref{defft}) 
is different in even and odd dimensions. 

For  $d+1$ even: the contour is given by 
\begin{eqnarray}\label{muo}
\mu_0( \lambda) &=& \frac{1}{2}\left(  \int_{ R+ i \delta}  + \int_{R- i \delta} \right) 
\frac{ du}{2\pi} e^{- i \lambda u}  W_0(u), \qquad\qquad d+1  \; \hbox{even}.
\end{eqnarray}
The contour is the sum of the line integrals above and below the real line. 
Since $W_0(u)$ has no branch cuts on the real line, the two line integrals can be 
replaced by a single line integral above the real line as shown in figure \ref{fig4}. 
We call this contour $C_e$. 
The 
normalization  of $W_0$   is such that   results in the factor $N(d)$ in (\ref{defnd})  for 
$d+1$ even in $\mu_0(\lambda)$ \footnote{Our normalization of $W_0(u)$ differs from that in \cite{Sun:2020ame}.}.

For $d+1$ odd, the contour for obtaining the Plancherel measure from its 
Fourier transform is given by 
\begin{eqnarray}\label{muo1}
\mu_0(\lambda) = -\frac{i}{\pi} \log(\tilde R) 
   \left(  \int_{ R- i \delta}  - \int_{R+i \delta} \right)   \frac{du}{2\pi}  e^{- i \lambda u} W_0(u).
\end{eqnarray}
Since $W_0(u)$ has no branch cuts on the real line,  the contour can be deformed to 
a small circle around the origin as shown in figure \ref{fig5}.  We call this contour $C_o$. 
Again, the normalization of $W_0$ and the factors in the equation (\ref{muo1}) are such that 
results in the factor $N(d)$  in (\ref{defnd}) for $d+1$ odd.

 We can now substitute 
 $W_0$ in the RHS of (\ref{diffrel}) we obtain 
\begin{equation}
\frac{d^2}{d u^2}W_p(u)-(p-\frac{d}{2})^2W_p(u) =  \frac{ (d+2)!}{ p! ( d- p)!} \frac{ ( 1+ e^{-u} ) 
e^{- \frac{ ( d+ 2) }{2} u } }{ ( 1- e^{-u} )^{ d+3}}.
\end{equation}
For $p\neq \frac{d}{2}$, the general solution of the differential equation is given by 
by the following  function depending on constants $c_1, c_2$. 
\begin{eqnarray}
   &&F(x) =c_1x^{\frac{d}{2}-p}+c_2x^{p-\frac{d}{2}} +  \\ \nonumber
  && \frac{x^{-\frac{d}{2}-p} \Gamma (d+3) }{(d-2p){\Gamma (p+1) \Gamma (d-p+1)}}
    \bigg\{
    x^{2 p} [B(x;  d-p+1,-d-1)-2 B(x; d-p+1,-d-2)]   \\ \nonumber
    & &\qquad\qquad\qquad\qquad\qquad\qquad   +x^d [2 B(x;  p+1,-d-2)-B(x_; p+1,-d-1)] \bigg\}.
\end{eqnarray}
To fix the integration constants, let us first expand in small $x$, we get 
\begin{eqnarray}
     F(x) =c_1 x^{p-\frac{d}{2}}+\frac{c_2 x^{\frac{1}{2} (d-2 p)}}{d-2 p}+\cdots
 \end{eqnarray}
To fix the boundary conditions, we demand that  $W_p(u)$ is analytic in $p$. Therefore
we demand that 
at $p=0$, the function obeys  the  expansion obeyed by 
$W_0$  given in (\ref{muo}). 
This   implies that we set  $c_1 =0$.  Further more note that $W_0$ has a zero at $x=-1$, this must 
be true for $W_p$ as well, since the factor $ \frac{  1+x }{ 1-x }$  arises due to the kinematic factor that the 
we are examining partition functions for bosons \footnote{The partition function of 
the  simple harmonic oscillator also admits such a factor \cite{Anninos:2020hfj}. }. 
Thus, setting $c_1 =0$ and demanding that $W_p(x)$ admits a zero at $x=-1$ we obtain 
\begin{eqnarray}
     c_2&= &\frac{\Gamma (d+3)}{\Gamma (p+1) \Gamma (d-p+1)}   \\ \nonumber
     && \times   \bigg\{ 2 (-1)^{2 p-d} B(-1; d-p+1,-d-2)+(-1)^{2 p-d+1} B(-1; d-p+1,-d-1) \\  \nonumber
     & & -2  B (-1; p+1,-d-2)+  B(-1; p+1,-d-1)\bigg\}.
\end{eqnarray}
Imposing these boundary conditions lead to the following expression for $W_p(x) $
\begin{eqnarray}
     W_p(x)&=&\frac{x^{-\frac{d}{2}-p}\Gamma (d+3)}{(d-2 p) \Gamma (p+1) \Gamma (d-p+1)}\bigg\{
     x^{2 p} \big[ B(x; d-p+1,-d-1)-2 B(x; d-p+1,-d-2) \big] \nonumber\\
     &&+2 (-1)^{2 p-d} x^d B(-1; d-p+1,-d-2)+(-1)^{-d+2 p+1} x^d B(-1; d-p+1,-d-1)\nonumber\\
     &&+x^d \big[ 2 B(x; p+1,-d-2)-B(x; p+1,-d-1)\nonumber\\
     &&-2 B(-1; p+1,-d-2)+B(-1; p+1,-d-1) \big] \bigg\}.
\end{eqnarray}
Though this  expression for $W_p(x)$ seems  non-illuminating, 
 it can be written in terms of $AdS$ Harish-Chandra characters as follows
\begin{eqnarray} \label{niceform}
     W_p(u)e^{( -\frac{d}{2} + p ) u}&=&\frac{1+e^{-u}}{1-e^{-u}}\sum_{i=0}^{p}(-1)^i\binom{d-2i}{p-i}\frac{e^{-(d-p-i)u}}{(1-e^{-u})^{d-2i}}, \\ \nonumber
     &=& \frac{1+e^{-u}}{1-e^{-u}}\sum_{i=0}^p \chi_{( d-2i,  p -i)}^{AdS}, \\ \nonumber
     \chi_{( d-2i,  p -i)}^{AdS} &=&  \binom{d-2i}{p-i}\frac{e^{-(d-p-i)u}}{(1-e^{-u})^{d-2i}}.
\end{eqnarray}
We have verified the above identity for all $p$-forms in  dimensions  $   2\leq d \leq 20$. 
We would like to emphasise that 
as a cross check we have also directly  verified that $W_p(u)$ given in (\ref{niceform}) solves the 
differential equation for $W_p(u)$  in (\ref{diffrel}). 
In the appendix  \ref{contourpes} for $ d+1$ even, we have performed the Fourier transform in (\ref{invft}) 
using the method of residues 
 and have demonstrated that the  the result for $W_p(u)$ is as given in  (\ref{niceform}). 
Therefore the method of residues 
provides another cross check especially for the choice of the boundary conditions we used in the 
differential equation (\ref{diffrel}) 
 to obtain $W_p(u)$.

Let us now examine the case $p = \frac{d}{2}$, this situation occurs 
only when $d+1$ is odd.  This case needs a separate discussion one of the independent solution
becomes logarithmic. 
For this case the differential equation  determining $W_p(u)$ 
reduces to 
\begin{align} \label{pddiff}
     \frac{d^2}{d u^2}W_p(u)=\frac{\Gamma (2 p+3)\left(e^{-u}+1\right) \left(e^{-(p+1) u}\right)}{\Gamma (p+1)^2\left(1-e^{-u}\right)^{2 p+3}}, \qquad\qquad\qquad p = \frac{d}{2} .
\end{align}
The  most general solution of this differential equation is given by 
\begin{equation}
    F(x) =c_1 \log (x)+c_2 +\frac{x^{p+1} \Gamma (2 p+3) \, _2\tilde{F}_1(p+1,2 p+2;p+2;x)}{\Gamma (p+2)},
\end{equation}
where $x=e^{-u}$ and 

${}_2\tilde{F}_1(p+1,2 p+2;p+2;x)$  is the regularised hypergeometric function defined 
in (\ref{defhyper}). 
It is clear that we need to set $c_1 =0$ if we wish to obtain an 
expression in terms of Harish-Chandra characters. 
Again demanding that $F(x)$ has a zero at $x=-1$, we obtain 
\begin{eqnarray}
    c_2&=&\frac{(-1)^{p} \Gamma (2 p+3)}{\Gamma (p+2)}\; {}_2\tilde{F}_1(p+1,2 p+2;p+2;-1) , \\  \nonumber
    &=& (-1)^p.
    \end{eqnarray}

Substituting these boundary conditions, the solution of the differential equation for 
$p=\frac{d}{2}$ is given by 
\begin{align}
    W_{p=\frac{d}{2}}(x)=\frac{ \Gamma (2 p+3)}{\Gamma (p+2)} 
     \,  x^{p+1}{} _2\tilde{F}_1(p+1,2 p+2;p+2;x) +(-1)^p, \qquad\qquad p = \frac{d}{2} 
\end{align}
Again it can be verified that the above expression can be written in terms of Harish-Chandra characters as follows
\begin{equation} \label{chrpd}
    W_{p=\frac{d}{2}}(u)
%    &=\frac{\left(e^{-u}\right)^{p+1} \Gamma (2 p+3) \, _2\tilde{F}_1\left(p+1,2 p+2;p+2;e^{-u}\right)}{\Gamma (p+2)}+(-1)^p\nonumber\\
    =\frac{1+e^{-u}}{1-e^{-u}}\sum_{i=0}^{p}(-1)^i \chi_{(d-2i, p -i) }^{AdS}, \qquad\qquad  p =\frac{d}{2} 
\end{equation}
where we have re-introduced $u = - \log x $. 
Note that in this case $p=\frac{d}{2}$, $\nu=\frac{d}{2}-p=0$.  
As a cross check we have also verified that the Fourier transform in (\ref{chrpd}) satisfies the 
differential equation (\ref{pddiff}). 

We conclude that for all value of $p, d$, the expression for the transform $W_p(u)$ is 
given in (\ref{niceform}).

\subsection*{The one loop determinant} 

Let us now finally use $W_p(u)$ in equation (\ref{gdet}) to write down the one loop determinant of the 
co-exact $p$-form on $AdS_{d+1}$.
For this we need to specify the contour $C$. 
We have already used the fact that our results should be continuous in $p$. 
Therefore we choose the 
contour to be the same as the one used for the case $p=0$ in (\ref{muo}) and (\ref{muo1}) for $d+1$ even 
and $d+1$ odd respectively.

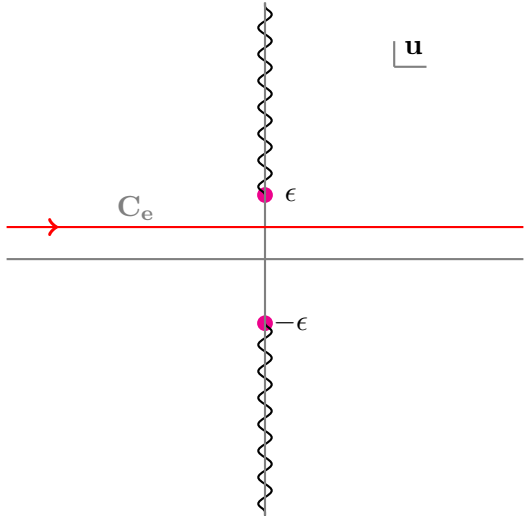
\begin{figure}[h]
\centering
\begin{tikzpicture}[thick,scale=0.85]
\filldraw[magenta] 
                (0,1) circle[radius=3pt]
                (0,-1) circle[radius=3pt];
\draw [decorate,decoration=snake] (0,-1) -- (0,-4);
\draw [decorate,decoration=snake] (0,1) -- (0,4);
%\draw (7.9,0) node {$\Re(w)$};
%\draw (0,6.5) node {$\Im(w)$};
%\draw (5.5,2.5) node {$\Re(w)>0$};
%\draw (-5.5,2.5) node{$\Re(w)<0$};
%\draw (3,-3) node{$|w|=1$};
\draw[gray,thick] (-2,0.8) node{$\mathbf{C_e}$};
\draw (0.4,1) node{$\mathbf{\epsilon}$};
\draw (0.4,-1) node{$\mathbf{-\epsilon}$};
\draw[gray, thick] (0,0) -- (0,4);
\draw[gray, thick] (0,0) -- (0,-4);
\draw[gray, thick] (0,0) -- (4,0);
\draw[gray, thick] (0,0) -- (-4,0);
\draw
[
postaction={decorate,decoration={markings , 
mark=at position 0.20 with {\arrow[red,line width=0.5mm]{>};}}}
][red, thick] (-4,0.5)--(0,0.5);
\draw[red, thick] (0,0.5)--(4,0.5);
\draw[gray, thick] (2,3) -- (2,3.4);
\draw[gray, thick] (2,3) -- (2.5,3);
\draw (2.3,3.3) node{$\mathbf{u}$};
\end{tikzpicture}
\caption{Contour $C_e$ in the $u$-plane for even $AdS_{d+1}$} \label{fig4}
\end{figure}

For even $d+1$  substituting (\ref{niceform}) in (\ref{gdet})  and using the contour in (\ref{muo}) 
we obtain
\begin{eqnarray}
-\frac{1}{2}\log(  {\rm det}_T \Delta_p^{AdS_{d+1} } )  = 
\frac{1}{2} \left( \int_{R+ i \delta} + \int_{ R- i \delta} \right) 
  \frac{du}{ 4 \sqrt{ u^2 + \epsilon^2}} e^{  ( - \frac{d}{2} +p ) \sqrt{ u^2 + \epsilon^2}  }  W_p(u) , 
\\ \nonumber
    W_p(u)e^{( -\frac{d}{2} + p ) u}=\frac{1+e^{-u}}{1-e^{-u}}\sum_{i=0}^{p}(-1)^i\binom{d-2i}{p-i}\frac{e^{-(d-p-i)u}}{(1-e^{-u})^{d-2i}}, \qquad\qquad d+1 \; \hbox{even} .
\end{eqnarray}
Since there is no branch cut on the real line, 
the contour below the real line and above the real line are equivalent. The contour is shown in 
figure \ref{fig4}. 
Furthermore note that $W_p( u) = W_p(-u)$ when $d+1$ is even. 
The remaining terms  in the integrand are symmetric in $u$.  
Thus we can restrict the integral over the  positive real axis.  Using these inputs and 
finally taking the $\epsilon\rightarrow 0$ limit, we obtain 
\begin{eqnarray}\label{eads}
-\frac{1}{2}\log(  {\rm det}_T \Delta_p^{AdS_{d+1} } )  = \int_0^\infty \frac{du}{2u} 
\frac{1+e^{-u}}{1-e^{-u}} \sum_{i =0}^p (-1)^i \chi_{( d-2i, p -i ) }^{AdS} (u) , 
\qquad \qquad   d+1 \; \hbox{even}. \nonumber \\
\end{eqnarray}

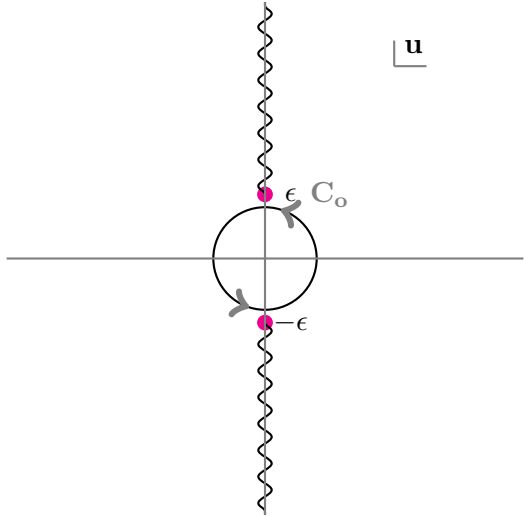
\begin{figure}[h]
\centering
\begin{tikzpicture}[thick,scale=0.85]
\filldraw[magenta] 
                (0,1) circle[radius=3pt]
                (0,-1) circle[radius=3pt];
\draw [decorate,decoration=snake] (0,-1) -- (0,-4);
\draw [decorate,decoration=snake] (0,1) -- (0,4);
\draw
[
postaction={decorate,decoration={markings , 
mark=at position 0.20 with {\arrow[gray,line width=1mm]{>};}}}
]
[
postaction={decorate,decoration={markings , 
mark=at position 0.70 with {\arrow[gray,line width=1mm]{>};}}}
]
(0,0) circle[radius=0.8cm];
%\draw (7.9,0) node {$\Re(w)$};
%\draw (0,6.5) node {$\Im(w)$};
%\draw (5.5,2.5) node {$\Re(w)>0$};
%\draw (-5.5,2.5) node{$\Re(w)<0$};
%\draw (3,-3) node{$|w|=1$};
\draw [gray,thick](1,1) node{$\mathbf{C_o}$};
\draw (0.4,1) node{$\mathbf{\epsilon}$};
\draw (0.4,-1) node{$\mathbf{-\epsilon}$};
\draw[gray, thick] (0,0) -- (0,4);
\draw[gray, thick] (0,0) -- (0,-4);
\draw[gray, thick] (0,0) -- (4,0);
\draw[gray, thick] (0,0) -- (-4,0);
\draw[gray, thick] (2,3) -- (2,3.4);
\draw[gray, thick] (2,3) -- (2.5,3);
\draw (2.3,3.3) node{$\mathbf{u}$};
\end{tikzpicture}
\caption{ Contour $C_o$ in the $u$-plane for even $AdS_{d+1}$} \label{fig5}
\end{figure}

For $d+1$ odd,  we substitute (\ref{niceform}) in (\ref{gdet}) and use the contour (\ref{muo}). 
This results in  the following 
\begin{eqnarray}
-\frac{1}{2}\log(  {\rm det}_T \Delta_p^{AdS_{d+1} } ) &=& 
\frac{1}{2\pi i } \log(\tilde R) 
   \left(  \int_{ R- i \delta}  - \int_{R+i \delta} \right)   \frac{du}{ 2 \sqrt{ u^2 + \epsilon^2}} e^{  ( - \frac{d}{2} +p ) \sqrt{ u^2 + \epsilon^2}  }  W_p(u) , \nonumber \\
   & & \nonumber \\
   & & \qquad\qquad d+1 \; \hbox{odd} .
\end{eqnarray}
The branch cut in the integrand occurs only on the imaginary axis, choosing 
chosing $\delta <<\epsilon$, we can deform the contour to  a small circle around $u=0$ and then 
take $\epsilon\rightarrow 0$.  The contour is shown in figure \ref{fig5}. 
Therefore we obtain 
\begin{eqnarray} \label{oads}
-\frac{1}{2}\log(  {\rm det}_T \Delta_p^{AdS_{d+1} } ) = 
  \frac{1}{2\pi i }  \log(\tilde R) \int_{C_0} \frac{du}{2u} 
  \frac{1+e^{-u}}{1-e^{-u}} \sum_{i =0}^p (-1)^i \chi_{( d-2i, p -i ) }^{AdS} (u) ,
  \qquad  d+1 \; \hbox{odd} .\nonumber \\
  \end{eqnarray}
  
  In the next section we will verify that (\ref{eads}) is indeed the expression for the one loop partition 
  of  all $p$-forms 
  in even dimensional $AdS$ space 
by comparing it  to corresponding partition function on even dimensional spheres. 
  For the odd dimensions we will provide a cross check that  (\ref{oads}) is the partition function for 
      $p = \frac{d}{2}$ in section \ref{pfhypcy}.

\subsection{Trace anomaly cross check for even $AdS$}\label{section2.2}

In conformally flat backgrounds and in even dimensions,  free energies  are proportional to the 
trace anomaly coefficient $a_{d+1}$, that appears with the Euler density as defined by the 
expectation value of the trace of the  stress tensor 
\begin{eqnarray}
\langle T^\mu_\mu \rangle = \frac{1}{ (4\pi )^{\frac{d+1}{2}}} 
\left( \sum_j c_{(d+1) j}  I_{j+1}^{(d+1)}  - ( -1)^{ \frac{ ( d+ 1)}{2} } a_{d+1} E_{d+1} \right) .
\end{eqnarray}
Here $E_{d+1}$ is the Euler density which is given by 
\begin{eqnarray} \label{euden}
E_{d+1}  = \frac{1}{ 2^{ \frac{d+1}{2}}} \delta^{\nu_1\cdots \nu_{d+1}}_{\mu_1\cdots \mu_d} 
R^{\mu_1\mu_2}_{\;\;\;\nu_1\nu_2} \cdots R^{\mu_d \mu_{d+1}}_{\;\;\;\nu_d \nu_{d+1}} 
\end{eqnarray}
 and $I_j^{(d+1)}$ are independent Weyl invariants of weight 
 $- (d+1)$. It is known that for instance in $d+1=4$ dimensions, the variation of the partition function 
in conformally flat back ground with respect to the metric is given by \cite{Brown:1977sj,Herzog:2013ed}
\begin{eqnarray}
\frac{2}{ \sqrt{g} } \frac{\delta \log {\cal Z}}{ \delta g_{\mu\nu} } &=& 
\langle T^{\mu\nu} \rangle , \\ \nonumber
&=& -\frac{a_4}{(4\pi)^2}\left(g^{\mu\nu}(\frac{R^2}{2}-R_{\lambda\rho}^2)+2R^{\mu\lambda}R^{\nu}_{~\lambda}-\frac{4}{3}RR^{\mu\nu}\right).
\end{eqnarray}
This equation is true for both $S^4$ and $AdS_4$, both these spaces are conformally flat. 
Integrating the above equation, we obtain 
\begin{equation}
\log {\cal Z} = a_4  \int d^4 x \sqrt{g} S ( R^{(2)} ) .
\end{equation}
Here $S(R^{(2)} )$, refers to a  function  made of  the various   quadratic  invariants of the curvature tensor. 
Now the curvature tensor  on spheres and anti-de Sitter spaces  of unit radius  differ 
just by a sign. 
\begin{equation}\label{curvsig}
R_{\mu\nu\rho\sigma}|_{S^{d+1} }  = g_{\mu\rho} g_{\nu\sigma} - g_{\mu\sigma} g_{\nu\rho} ,  \qquad\qquad
R_{\mu\nu\rho\sigma}|_{AdS_{d+1}}  =  g_{\mu\sigma} g_{\nu\rho}  - g_{\mu\rho} g_{\nu\sigma} .
\end{equation}
This implies that   $S(R^{(2)} )$ evaluated on $AdS_4$ and $S^4$ would be identical. 
Furthermore since both the spaces are Einstein, we see that  $S(R^{(2)})$ would 
be constant in the space.  This leads us to conclude
\begin{eqnarray}
\frac{ \log {\cal Z}[AdS_4]}{ \log{\cal Z} [S^4]} = 
\frac{ {\rm Vol } ( AdS_4) } { {\rm Vol } ( S^4)} 
&=& \frac{1}{2} .
\end{eqnarray}
To obtain the last line we have substituted the regularised volume of $AdS^4$ given 
in (\ref{regvol}) and the volume of sphere given by 
\begin{eqnarray}
{\rm Vol} ( S^{d+1})  = 2 \frac{\pi^{\frac{d}{2} +1 } }{ \Gamma ( \frac{d}{2} +1) },
\end{eqnarray}
which results in 
\begin{equation}\label{ratiov}
\frac{{ \rm Vol } (AdS_{d+1}) } { {\rm Vol }(S^{d+1}) } 
     = \frac{  (-1)^{\frac{d+1}{2}} }{2} .
\end{equation}
We can repeat this analysis  in $d+1=6$. In this case, the equation corresponding to (\ref{curvsig}) is known 
and is given by  \cite{Herzog:2013ed}
\begin{eqnarray}
\frac{2}{ \sqrt{g} } \frac{\delta \log {\cal Z}}{ \delta g_{\mu\nu} } &=& 
\langle T^{\mu\nu} \rangle , \\ \nonumber
&=& - \frac{a_6}{( 4\pi)^3} \left[ \frac{3}{2} R^\mu_\lambda R^\nu_\sigma R^{\lambda\sigma}
- \frac{3}{4} R^{\mu\nu} R^\lambda_\sigma R^\sigma_\lambda - \frac{1}{2} g^{\mu\nu} 
R^{\sigma}_{\lambda} R^\lambda_\rho R^\rho_\sigma, \right. 
 \\ \nonumber
& &  \left. 
- \frac{21}{20} R^{\mu\lambda} R^\nu_\lambda R + \frac{21}{40} g^{\mu\nu} R^\sigma_\lambda R^\lambda_\sigma R + \frac{39}{100} R^{\mu\nu} R^2 - \frac{1}{10} g^{\mu\nu} R^3 \right].
\end{eqnarray}
Integrating this equation we get 
\begin{equation}\label{ads6p}
\log {\cal Z} = a_6  \int d^6 x \sqrt{g} S ( R^{(3)} ) ,
\end{equation}
where the  $S( R^{(3)}) $ is a function of cubic invariants constructed out of the curvature tensor. 
Using the fact that the curvature tensor on $AdS$ and the sphere differ by a sign as in (\ref{curvsig}), 
we see that 
\begin{eqnarray}\label{act6}
S( R^{(3)} )|_{AdS_6} = - S( R^{(3)} )|_{S^6} .
\end{eqnarray}
Using  (\ref{ads6p}), (\ref{ratiov}) and  (\ref{act6})  we conclude that 
\begin{eqnarray}
\frac{ \log {\cal Z}[AdS_6 ]}{ \log{\cal Z} [S^6]} &=& 
-\frac{ {\rm Vol } ( AdS_6) } { {\rm Vol } ( S^6)} 
= \frac{1}{2} .
\end{eqnarray}
Indeed noting the fact that the Euler density in (\ref{euden}) is a  homogenous  polynomial  composed 
of curvature tensors of order 
$\frac{d+1}{2}$ allows us to conclude that in $d+1$ dimensions that the free energies are 
given by 
\begin{equation}\label{adsad}
\log {\cal Z} = a_{d+1}   \int d^{d+1}  x \sqrt{g} S ( R^{(\frac{d+1}{2} )} ) .
\end{equation}
Here $S ( R^{(\frac{d+1}{2} )} ) $ is a function  invariants made of $\frac{d+1}{2}$ powers of 
the curvature tensor. 
From the properties of the curvature in $AdS$ and spheres we have 
\begin{eqnarray}\label{adsad1}
S( R^{(\frac{d+1}{2} )} )|_{AdS_{d+1} } =  (-1)^{\frac{d+1}{2}} S( R^{(\frac{d+1}{2} )} )|_{S^{d+1} }.
\end{eqnarray}
Then from (\ref{adsad}), (\ref{ratiov}) and (\ref{adsad1}) we conclude
\begin{eqnarray} \label{predict}
\frac{ \log {\cal Z}[AdS_{d+1}] } { \log{\cal Z} [S^{d+1} ]} &=& 
(-1)^{d+1}  \frac{ {\rm Vol } ( AdS_{d+1} ) } { {\rm Vol } ( S^{d+1} )} 
= \frac{1}{2} .
\end{eqnarray}

Thus from the trace anomaly we  can conclude that free energies in two conformally flat backgrounds 
 is the same functional of the curvatures.  This leads to the prediction that the 
 ratio of free energies in even dimensional $AdS$ and spheres is half. 
 Since we have written down the partition function in both these spaces in 
 terms of Harish-Chandra characters  we can verify that this prediction 
 is indeed  true.
 
  Let us examine the coefficient of the logarithmic divergence 
of the partition function. This is the only term which does not depend 
 any prescription to regulate the integral in (\ref{eads}). 
The logarithmic divergence is determined by the coefficient of the $\frac{1}{u}$ term of 
the integrand. 
This can be extracted by considering a small circle $C_r$  of radius  $r$ around the origin.
Therefore the coefficient of the logarithmic divergence  of the partition function of the 
co-exact $p$-form  in even $AdS_{d+1}$ given in
 (\ref{eads}) 
 \begin{eqnarray}\label{eads2}
&&-\frac{1}{2}\log(  {\rm det}_T \Delta_p^{AdS_{d+1} } )|_{\rm log \; divergence}   
= \\ \nonumber
&& \qquad\qquad\qquad \frac{1}{2\pi i } \int_{C_r}  \frac{du}{2u} 
\frac{1+e^{-u}}{1-e^{-u}} \sum_{i =0}^p (-1)^i  \binom{d-2i}{p-i}\frac{e^{-u(d-p-i)} }{(1-e^{-u} )^{d-2i} } \\ \nonumber
&&  \qquad\qquad\qquad = \frac{1}{2\pi i } \int_{C_r}  \frac{du}{2u} 
\frac{1+e^{u}}{1-e^{u}} \sum_{i =0}^p (-1)^i  \binom{d-2i}{p-i}\frac{e^{u( p-i)}}{(1-e^{u} )^{d-2i}} .
\end{eqnarray}
In the last line we have also used the fact that $d+1$ is even. 
Note that in this integration $u = r^{i \theta}$, and  $\theta$ runs from $0$ to $2\pi$.
The result is invariant if we start the contour at  $\theta$  from $\pi$ and take it all the way to $3\pi$
in the counter-clock wise direction.
The contour still remains the same. This effectively is a change of variable $\theta\rightarrow \hat \theta + \pi$, 
where $\hat \theta$ runs from $0$ to $2\pi$. 
But  performing this change of variables sends  $u \rightarrow -u$. 
Therefore  we see that 
\begin{eqnarray}\label{eads3}
&& -\frac{1}{2}\log(  {\rm det}_T \Delta_p^{AdS_{d+1} } )|_{\rm log\; divergence}    \\ \nonumber
&& \qquad\qquad\qquad  = \frac{1}{2\pi i } \int_{C_r}  \frac{du}{2u} 
\frac{1+e^{-u}}{1-e^{-u}} \sum_{i =0}^p (-1)^i  \binom{d-2i}{p-i}\frac{e^{-u( p-i)}}{(1-e^{-u} )^{d-2i}} .
\end{eqnarray}
As a cross check we have explicitly  verified that indeed that coefficient of the $\frac{1}{u}$ term 
in (\ref{eads3}) agrees with that of (\ref{eads2}). 
Now adding (\ref{eads2}) and (\ref{eads3}) we see that 
\begin{eqnarray}
-\frac{1}{2}\log(  {\rm det}_T \Delta_p^{AdS_{d+1} } )|_{\rm log\; divergence}   
&=& \frac{1}{2}  \frac{1}{2\pi i } \int_{C_r}  \frac{du}{2u} \frac{1+e^{-u}}{1-e^{-u}} 
\left(  \sum_{i =0}^p (-1)^i   \chi^{dS}_{ d-2i, p-i} (u)  \right),  \nonumber \\
&=&- \frac{1}{4} \log(  {\rm det}_T \Delta_p^{S^{d+1} } )|_{\rm log\; divergence} .  
\end{eqnarray}
In the first line of the above equation we have used the definition of the  
$SO(1, d+1)$ Harish-Chandra character 
 given in (\ref{defchr}). 
We have shown that 
the coefficient of the logarithmic divergence in the partition function of the 
 co-exact $p$-form on even dimensional $AdS_{d+1}$ is 
half that on the even sphere $S^{d+1}$. Therefore
it is clear that the same fact holds for the logarithmic divergence 
of the partition function of the $p$-forms in these spaces since the co-exact forms form the key ingredient of the 
partitions functions as given by 
(\ref{gfpform}) and (\ref{p-fads}). 
 This concludes our check of prediction  given in  (\ref{predict}).

 \section{Hyperbolic cylinders and branched spheres} \label{section3}

 In this section we study  one loop partition functions of 
 conformal scalars as well as conformal $p$-forms on  hyperbolic cylinders. We show these 
 partition functions can be written 
 in terms of Harish-Chandra characters. 
 Partition functions on hyperbolic cylinders naturally arise on evaluating entanglement entropy 
 of  conformal field theories across a spherical entangling surface \cite{Casini:2011kv}. 
When the ratio of the radius of $S^1$ to that of $AdS$ is $q$ the cylinder is referred to  as $S^1_q \times AdS_d$.
Then 
the R\'{e}nyi entropy 
across a spherical entangling surface can be evaluated given the 
 partition function of a conformal field theory on the hyperbolic cylinder. 
  The work of \cite{Anninos:2020hfj,Sun:2020ame} and the previous sections 
  in this paper have demonstrated that one loop partition 
 functions on spheres and anti-de Sitters spaces have nice character integral 
 representations. 
 It is natural to ask the question whether the same can be said about the  partition 
 functions on  hyperbolic cylinders.

Hyperbolic  cylinders $S^1_q\times AdS_d$ are known to be conformally equivalent to 
branched spheres  $S_q^{d+1}$ \cite{Casini:2011kv}. 
Therefore we expect that the character representation 
of the partition function 
of conformal scalars on hyperbolic cylinders to agree with that of the branched sphere. 
In section  \ref{confscal}  we indeed show that this expectation is true. 
In section  \ref{pfhypcy} we show that the character integral representation of 
conformal $p$-form partition functions on 
the hyperbolic cylinder agrees with only the bulk character  of the corresponding partition 
function on the sphere.  This is consistent with the earlier observations of 
\cite{Casini:2013rba,Huang:2014pfa,Donnelly:2014fua,Donnelly:2015hxa,Casini:2015dsg,David:2020mls} that partition function on hyperbolic cylinders  miss out the 
edge modes or the non-extractable classical contribution to entanglement entropy. 
The character integral representation of the conformal $p$-form we derive 
generalises the observation seen first for $1$-forms, to conformal  $p$-forms in arbitrary 
dimensions. 
Finally in section \ref{branch},  we obtain a character integral representation for 
$1$-forms in arbitrary dimensions on branched spheres. 
Using this input, together with the 
observations for character representation of conformal $p$-forms on hyperbolic 
cylinders $S^1_q\times AdS_d$ we propose  the character integral representation of the partition 
functions of 
co-exact $p$-form on branched spheres in all dimensions. 
We verify that this proposal agrees with previous evaluations of these partition functions by
\cite{Dowker:2017flz}.

     \subsection{Conformal scalars} \label{confscal}

  We will first  evaluate the partition function of conformal  scalars on the hyperbolic cylinder and show 
  that it can be written in terms of Harish-Chandra characters. 
 We will see  these characters turn out to be characters for the sphere and 
  the partition function is identical to that of conformal scalars on the branched sphere. 
  
  \subsubsection*{Conformal scalars on $S^1_q\times AdS_{d}$}
  
The Weyl invariant action of the real scalar  in $d+1$ dimensions is given by 
\begin{equation}\label{action}
S =- \frac{1}{2}  \int d^{d+1} x \sqrt{g} (  \partial_\mu\phi \partial^\mu \phi   + \frac{d-1}{4 d} R \phi^2) .
\end{equation}
Here $R$ is the curvature. 
The metric on $S^1_q\times AdS_{d}$  is given by 
\begin{equation}
ds^2_{S^1_q\times AdS_{d}}  =  d\tau^2 +  du^2 + \sinh^2  u d\Omega^2_{d-1},
\end{equation}
where $\tau $ is the coordinate on the circle with the  identification $\tau \sim \tau + 2\pi q$. The radial 
coordinate on $AdS_d$ is $u$  and 
$\Omega_{d-1}$ refers to the $d-1$ sphere. 
Using this metric, we can evaluate  the curvature scalar on the hyperbolic cylinder which is given by 
\begin{eqnarray}
R|_{S^1_q\times AdS_{d} } = - d( d-1) .
\end{eqnarray}
The partition function of Weyl invariant scalar on this background  is given by the determinant
\begin{align}
    \mathcal{Z}[S^1_q\times AdS_{d}]
    &=\left(\frac{1}{\det(-\partial_{\tau}^2-\Delta_0+m_{S^1_q \times AdS_d}^{2} )}\right)^{\frac{1}{2}},
\end{align}
where the mass  arises from the curvature coupling and is given by 
\begin{equation}\label{massads}
m_{S^1_q\times AdS_{d} }^2    =    - ( \frac{d-1}{2} )^2 .
\end{equation}
Here $\Delta_{(0)}$ is the spin-0 Laplacian on $AdS_d$. 
We  decompose the scalar using the eigen modes of the spin-0 Laplacian on
$AdS_d$ and the Kaluza-Klein modes on the circle $S^1$. 
The eigen values of  the spin-0 Laplacian  on $AdS_d$  are given by 
\begin{equation}\label{eigen}
 \Delta_{(0)} \psi^{\{\lambda, u\}} = -\left[  \lambda^2 + (  \frac{d-1}{2} )^2 \right] 
 \psi_\lambda^{\{\lambda, u\}} ,
 \end{equation}
 $\psi^{ \{\lambda, u\} }$  are the corresponding eigen functions, $\{u\}$ labels other quantum numbers
 on $AdS_d$. 
Using these eigen values  and the Kaluza-Klein decomposition of the partition function. 
we obtain
\begin{eqnarray}\label{feq1}
\log \mathcal{Z}[S^1_q\times AdS_{d}]
  %-\frac{1}{2}\log(\det(-\partial_{\tau}^2-\Delta_0)-m_0^2)
   =
 -\frac{1}{4}  \sum_{n=-\infty}^\infty \int_{-\infty}^\infty d\lambda \mu_{0}^{(d)}(\lambda) 
    \log \left( \frac{n^2}{q^2} +\lambda^2 
 \right).
\end{eqnarray}
Note that the shift in the eigen value of the Laplacian in (\ref{eigen}) precisely cancels the 
mass due to the curvature coupling. The mass in (\ref{massads}) 
saturates the Brietenholer-Freedman bound. 
We have labelled the Plancherel measure for scalars  with the superscript $(d)$ to indicate that 
we are in $AdS_d$. 
This measure is given by the expressions in (\ref{defmup}),  with $p=0$ and $d\rightarrow d-1$ since we
are in $AdS_d$. 
We replace the logarithm by the identity in (\ref{logiden})  to obtain
\begin{eqnarray}
\log \mathcal{Z}[S^1_q\times AdS_{d}]
 =  \frac{1}{4} \int_0^\infty \frac{d\tau}{\tau} \sum_{n=-\infty}^\infty 
  \int_{-\infty}^{\infty} d\lambda 
  \mu_{0}^{(d)}(\lambda)  ( e^{ -\tau( \lambda^2 + \frac{n^2}{q^2} ) } - e^{-\tau} ) .
\end{eqnarray}
Just as in the case of (\ref{sumpla}), by analytically continuing from large negative $d$ we have 
\begin{equation}
\int_0^\infty d\lambda \mu_0^{(d)} =0 .
\end{equation}
Therefore we can proceed by regulating the first term as 
\begin{equation}
\log \mathcal{Z}[S^1_q\times AdS_{d}]
  =
  \int_0^{\infty}\frac{d\tau}{4\tau}e^{-\frac{\epsilon^2}{4\tau}}\int_{-\infty}^{\infty} 
  d\lambda 
  \mu^{(d)}_0 (\lambda)  \left(  e^{ - \tau \lambda^2} + 2 \sum_{n=1}^\infty 
  e^{-\tau(\lambda^2+\frac{n^2}{q^2})  }\right)  .
\end{equation}
Now replace the Plancherel  measure  by its Fourier transform given by 
\begin{equation}
W_0^{(d)}( u) = \frac{1+ e^{-u} }{ 1- e^{-u}} \frac{ e^{-\frac{d-1}{2} u} }{ ( 1- e^{-u} )^{d-1}} .
\end{equation}
Note that here $d$  in equation (\ref{wo}) has been replaced by $d-1$  as we are in $AdS_d$. 
Then following  the same steps as in equations (\ref{man1}), (\ref{man11}) and (\ref{gdet}) 
we are led to 
\begin{eqnarray} \label{geoms}
\log \mathcal{Z}[S^1_q\times AdS_{d}]
  =\int_{C}
  \frac{du}{4\sqrt{\epsilon^2+u^2}}(1+2\sum_{n=1}^{\infty}e^{-\frac{n}{q}\sqrt{\epsilon^2+u^2}})W_{0}^{(d)} (u).
\end{eqnarray}
Here the contour is $C_o$ or $C_e$ as defined in figure \ref{fig4} and figure \ref{fig5}
 depending on whether $d$ is odd or even respectively. 
 
 For the case when $d$ is even substituting $W_0^{d}$  and using the 
 contour $C_e$ as shown in figure \ref{fig4} we obtain 
 \begin{equation}
 \log \mathcal{Z}[S^1_q\times AdS_{d}]
=
  \frac{1}{2} \int_{R + i \delta} \frac{du}{2u} 
  \frac{1+ e^{-u} }{ 1- e^{-u}} \frac{ e^{-\frac{d-1}{2} u} }{ ( 1- e^{-u} )^{d-1} } 
  \frac{ 1+ e^{ - \frac{1}{q} \sqrt{ \epsilon^2 + u^2} }}{  1- e^{ - \frac{1}{q} \sqrt{ \epsilon^2 + u^2} }}.
  \end{equation}
  We have summed the geometric series in (\ref{geoms}). 
  We can use the fact that the integrand is even to integrate only over the positive real axis, 
  and then take the $\epsilon\rightarrow 0$ limit and then take $\delta \rightarrow 0$. 
  Thus we arrive at 
  \begin{equation}\label{finres1ads}
  \log \mathcal{Z}[S^1_q\times AdS_{d}]
  =
  \int_0^\infty  \frac{du}{2u}  \frac{ 1+ e^{ - \frac{u}{q} } }{ 1- e^{ - \frac{u}{q}} } 
  \chi_{ (d, 0)\;{\rm conf}  }^{dS} ( u) ,
  \end{equation}
  where  the  $SO(1, d+1)$ Harish-Chandra character  is given by
  \begin{equation} \label{charconfsc}
   \chi_{(d, 0)\;{\rm conf}  }^{dS} ( u)  = 
   \frac{ e^{ - \frac{(d-1)}{2} t } + e^{ - \frac{d+1}{2} t } }{ ( 1- e^{-t} )^d } .
  \end{equation}
 We observe that this 
 character corresponds to that of the conformal scalar on $S^{d+1}$. \\
  The  $SO(1, d+1)$ Harish-Chandra character of the scalar of mass $m$ is given by 
  \begin{eqnarray} 
  \chi_{ (d, 0)\;{\nu }  }^{dS} ( u)  = 
     \frac{ e^{ -(  \frac{d}{2} +i \nu )  t } + e^{ -(  \frac{d}{2} - i \nu )  t } }{ ( 1- e^{-t} )^d } 
  \end{eqnarray}
  and $\nu$ is related to the mass by the following 
  \begin{eqnarray}\label{definu}
  i \nu =  \sqrt{  \frac{d^2}{4}  -   m^2}.
  \end{eqnarray}
  This mass induced by the curvature coupling on the sphere $S^{d+1}$  can be read out from 
  (\ref{action}) with $R = d( d+1)$.  We see that this mass is 
  \begin{eqnarray}\label{sphermass}
  m^2_{ S^{d+1}} = \frac{d^2 -1}{ 4} .
  \end{eqnarray} 
  Substituting this mass in  (\ref{definu}) we see that that the Harish-Chandra character reduces to that 
  seen in (\ref{finres1ads}).

  We can proceed with the same analysis for the case of $d$  is odd. The analysis is identical 
  except that the contour $C_e$ is replaced by $C_o$ of figure (\ref{fig5}). 
  The result is given by 
   \begin{equation}\label{finres1ads2}
    \log \mathcal{Z}[S^1_q\times AdS_{d}]
   =
  \frac{\log( \tilde R) }{2\pi i} \int_{C_o}  \frac{du}{2u}  \frac{ 1+ e^{ - \frac{u}{q} } }{ 1- e^{ - \frac{u}{q}} } 
  \chi_{ (d, 0)\;{\rm conf}  }^{dS} ( u) .
  \end{equation}
  
  The interesting thing to note from (\ref{finres1ads}) and (\ref{finres1ads2}) 
   is that we stared with the $SO(2, d-1)$ Harish-Chandra characters on 
  $AdS_d$. The sum over all the Kaluza-Klein modes resulted in  $SO(1, d+1)$ Harish-Chandra characters. 
  Note that the we have included the zero Kaluza-Klein mode as well.  In literature usually this is omitted 
  since it just results it is $q$ independent \cite{Hung:2014npa}. 
  However we see here that it is including this term, that the integrand organises as a character. 
    Furthermore observe that at $q=1$, these results are  
    identical to partition function of the conformal scalar  on sphere $S^{d+1}$. 
  Indeed, we will show that the partition function of conformal scalars on branched spheres 
  $S^{d+1}_q$ precisely agrees with that obtained   in (\ref{finres1ads}) and (\ref{finres1ads2}) 
  for $S^1_q\times AdS_d$.

 \subsubsection*{Conformal scalars on branched spheres $S_q^{d+1}$} 
 \label{conscalbrsp}
 
 The metric on the branched sphere is given by 
 \begin{equation}
 ds^2|_{S_q^{d+1}}  = \cos^2 \phi d \tau^2 + d\phi^2 + \sin^2 \phi d\Omega_{d-1}^2 ,
 \end{equation}
 where $\tau \sim \tau +2\pi q$ and $ 0\leq \phi \leq \frac{\pi}{2}$. 
 Given the  Weyl invariant action \eqref{action} and the Ricci curvature  $R=d(d+1)$, the 
 partition function of conformal scalar  on the branched sphere can be written as
 \begin{eqnarray}
     { \mathcal Z}[S_q^{d+1} ] =\frac{1}{ {\rm det} ( -\Delta_0+m^2_{S^{d+1}_q} )^{\frac{1}{2} } } .
 \end{eqnarray}
 The curvature induced mass  can be read off from  (\ref{sphermass}). 

 The eigenvalue and their corresponding  degeneracies  for  the scalar Laplacian on  the branched sphere
 are known \cite{DeNardo:1996kp}. They are labelled by $2$ integers
 \begin{equation}\label{eigenvalue}
     \lambda^{(0)} _{n,m} =(n+\frac{m}{q})(n+\frac{m}{q}+d), \qquad  n, m \in \{ 0, \cdots \infty \} 
 \end{equation}
 with degeneracies 
 \begin{eqnarray} \label{degeneracy}
         g^{(0)}_{n,m=0}&=&\binom{d+n-1}{d-1},    \qquad n  \in \{ 0, \cdots \infty \}  \\  \nonumber
         g^{(0)}_{n,m>0}&=&2\binom{d+n-1}{d-1}, \qquad  n  \in \{ 0, \cdots \infty \}, \qquad m \in \{ 1, \cdots \infty\} 
 \end{eqnarray}
 Therefore the free energy of the conformal scalar
 \begin{eqnarray}
    \log\mathcal{Z}[S^{d+1}_q] &=& - \frac{1}{2} 
    \sum_{n, m =0}^\infty  g_{n, m}^{(0)} \log \left( \lambda_{n, m}^{(0)}  + \frac{ d^2 - 1}{4} \right) ,
    \\ \nonumber
    &=& \int_0^\infty \frac{d\tau}{2\tau}  \sum_{n, m =0}^\infty
    g_{n, m }^{(0)} (  e^{ - \tau ( \lambda_{n, m}^{(0)} )}  - e^{-\tau} ) .
    \end{eqnarray}
    In the second line of the above equation we have again used the identity (\ref{logiden}). 
    It can be shown by looking at sufficiently large negative $d$ and using zeta function regularization
    for the summation over $m$ we obtain
    \begin{equation} 
    \sum_{n, m =0}^\infty g_{n, m }^{(0)}  =0.
    \end{equation} 
    Therefore we  can proceed by  regulating the first term 
    \begin{equation}
        \log\mathcal{Z}[S^{d+1}_q] =   \int_0^\infty \frac{d\tau}{2\tau} e^{-\frac{\epsilon^2}{4\tau} } 
       \sum_{n, m =0}^\infty g_{n, m }^{(0)} e^{ - \tau ( \lambda_{n, m}^{(0)} )}  .
    \end{equation}
    We can now  perform the Hubbard-Stratonovich trick  given in  (\ref{contint}) and  following the same
    steps as in equation (\ref{beginstep}) to (\ref{endstep}) we obtain 
    \begin{eqnarray} \label{spherman}
       \log\mathcal{Z}[S^{d+1}_q]  = \int_{\epsilon}^{\infty}\frac{dt}{2\sqrt{t^2-\epsilon^2}}\left(e^{i\nu\sqrt{t^2-\epsilon^2}}+e^{-i\nu\sqrt{t^2-\epsilon^2}}\right)f_q^{(0)} (i t).\
    \end{eqnarray}
    with $\nu = \frac{i}{2}$ and 
    \begin{eqnarray}
    f_q^{(0)} ( u ) &=& \sum_{n,m =0}^\infty g^{(0)}_{n,,m}e^{i(n+\frac{m}{q}+\frac{d}{2})u}\nonumber\\
    &=& \frac{e^{ \frac{i u d}{2}}}{ ( 1 - e^{i u } )^d } \frac{ 1 + e^{ i \frac{ u }{ q}} }{ 1 - e^{ i \frac{ u }{ q}}}.
    \end{eqnarray}
    Substituting these expressions in   \eqref{spherman} and taking the $\epsilon\rightarrow 0$ limit the partition function becomes
    \begin{equation} \label{branscalsp}
      \log\mathcal{Z}[S^{d+1}_q] = \int_0^\infty  \frac{dt}{2t}
      \frac{ 1+ e^{ -\frac{t}{q}} }{ 1- e^{ - \frac{t}{q} }}   \chi_{ (d, 0)\;{\rm conf}  }^{dS} ( u), 
    \end{equation}
    with the character as given in (\ref{charconfsc}). 
    
    For $d+1$ odd,  comparison with equation (\ref{finres1ads}),   with (\ref{branscalsp})
    we see that the integrands are identical. 
    For the case when $d+1$ is even  comparison of (\ref{finres1ads2})  with  (\ref{branscalsp})
   shows that the regularization independent quantity, the logarithmic 
    divergence of the expression in (\ref{branscalsp}) and that given in (\ref{finres1ads2}) 
    are identical provided we relate the 
    cut off to regulate the integral in 
     (\ref{branscalsp}) to $\tilde R$ in (\ref{finres1ads2}). 
    Indeed as recently mentioned in \cite{Nishioka:2021uef}, the agreement of  free energies of the conformal scalar on 
    the hyperbolic cylinder and the branched sphere coincides was verified till $d=100$ \footnote{See \cite{Sato:2021eqo} for a discussion for fermions.}. 
    It is interesting to note here that since  their 
    character integral representations coincide and therefore the agreement of the partitions functions
    is manifest.

    Given the partition function on the branched cylinder, one can evaluate the  universal contribution 
    to   R\'{e}nyi entropy across spherical entangling surfaces 
    in even dimensions from the logarithmic divergence of the 
    free energy ${\cal F}_q$ on hyperbolic cylinders.
    The R\'{e}nyi entropy $S_q$ and the entanglement entropy  $S_{\rm{EE}}$ are given by 
     \begin{eqnarray}
          S_q=\frac{-\mathcal{F}_q+q\mathcal{F}_{q=1}}{1-q}, \qquad\qquad
          S_{\rm{EE}}=\lim_{q\rightarrow 1}S_q.
    \end{eqnarray}
    In table \ref{table3}  we have listed both these entropies for the conformal scalar are listed 
    for even $4 \leq d+1 \leq 14$. These have been evaluated using the character integral representation
    We have seen that they precisely agree with earlier evaluations in \cite{Casini:2010kt}. 
    The IR finite term when $d+1$ is odd is also a regularization independent term. 
    We have evaluated this  finite part  from the character integral representation for the corresponding
    partition function 
     in table \ref{table4}. These precisely agree with that evaluated in table 1 of \cite{Klebanov:2011gs}.

    \begin{table}[ht]
\centering { \footnotesize{
\begin{tabular}{c|l|c}
\hline
$D $ & \quad $~~~~~~~~~~~~~~~~~~~~~~~-\mathcal{F}_q $ & $S_{\rm{EE}}$ \\
\hline 
& &  \\
$2$ & \quad$~~~~~~~~~~~~~~~~~~~~~~~\frac{q^2+1}{6 q} $ & $\frac{1}{3}$
   \\   
&  & \\
$4$ & \quad$~~~~~~~~~~~~~~~~~~~ -\frac{3 q^4+1}{360 q^3}$ & $ -\frac{1}{90}$  \\
& &  \\
$6$ & \quad$~~~~~~~~~~~~~~~~~~~~\frac{31 q^6+7 q^2+2}{30240 q^5}  $ &  $\frac{1}{756}$  \\
&  & \\
$8 $& \quad$~~~~~~~~~~~~-\frac{289 q^8+56 q^4+20 q^2+3}{1814400 q^7} $ &
$- \frac{23}{113400}$  \\
& &  \\
$10$& \quad $~~~~~~~~~\frac{6657 q^{10}+1188 q^6+462 q^4+99 q^2+10}{239500800 q^9}$
& $\frac{263}{7484400}$\\
& & \\ 
$12 $& \quad 
$-\frac{6803477 q^{12}+1153152 q^8+469040 q^6+117117 q^4+18200 q^2+1382}{1307674368000 q^{11}}$
&  $-\frac{133787}{20432412000} $\\
& &\\
$14 $& \quad 
$\frac{16018495 q^{14}+2620800 q^{10}+1095952 q^8+298155 q^6+56420 q^4+6910 q^2+420}{15692092416000 q^{13}}$
&  $\frac{157009}{122594472000} $\\
&  & \\
\hline
\end{tabular}
\caption{ Logarithmic divergence of the partition function of conformal scalars on branched spheres and 
universal terms in  entanglement entropies.  }
\label{table3}
}}
\end{table}

 \begin{table}[ht]
\centering { \footnotesize{
\begin{tabular}{c|l}
\hline
$d+1 $ &  ~~~~~~~~~~~~~~~~~~~~~~~ $-\mathcal{F}_{q=1}$ \\
\hline 
\\
$3$ & \quad$\frac{3 \zeta (3)-2 \pi ^2 \log (2)}{16 \pi ^2}$ 
   \\   
\\
$5$ & \quad$\frac{2 \pi ^2 \zeta (3)-15 \zeta (5)+2 \pi ^4 \log (2)}{256 \pi ^4}$  \\
 \\
$7$ & \quad$-\frac{82 \pi ^4 \zeta (3)-150 \pi ^2 \zeta (5)-945 \zeta (7)+60 \pi ^6 \log (2)}{61440 \pi ^6} $   \\
 \\
 $9$ & \quad$\frac{1588 \pi ^6 \zeta (3)-210 \pi ^4 \zeta (5)-13230 \pi ^2 \zeta (7)-26775 \zeta (9)+1050 \pi ^8 \log (2)}{6881280 \pi ^8} $   \\
 \\
 \\
  $11$ & \quad$-\frac{70146 \pi ^8 \zeta (3)+48500 \pi ^6 \zeta (5)-383670 \pi ^4 \zeta (7)-1338750 \pi ^2 \zeta (9)-1611225 \zeta (11)+44100 \pi ^{10} \log (2)}{1651507200 \pi ^{10}}$   \\
 \\
 \\
 $13$ & \quad$\frac{7157604 \pi ^{10} \zeta (3)+8436890 \pi ^8 \zeta (5)-24019380 \pi ^6 \zeta (7)-124289550 \pi ^4 \zeta (9)-248128650 \pi ^2 \zeta (11)-212837625 \zeta (13)+4365900 \pi ^{12} \log (2)}{871995801600 \pi ^{12}}$   \\
 \\
 \\
\hline
\end{tabular}}}
\caption{IR finite term or the `F'-term  on odd spheres for conformal scalars.} \label{table4}
\end{table}

\subsection{Conformal $p$-forms} \label{pfhypcy}

In \cite{David:2020mls} a procedure to fix gauge on hyperbolic cylinders was introduced. Using this method
the gauge invariant  partition  functions of the conformal $1$-form on $S^1_q\times AdS_3$ and 
the $2$-form on $S^1_q\times AdS_5$ was obtained. 
The conclusion in both the cases was the gauge invariant partition function can be thought of 
as the partition function of the tower of Kaluza-Klein tower of co-exact $p$-forms with the 
Hodge-de Rham Laplacian along the $AdS$ directions. 
Though this observation was explicitly demonstrated  only for the $1$-form in $d+1=4$ and $2$-form
in $d+1=6$, the method developed in \cite{David:2020mls} is such that the result can be 
extrapolated to conformal  $p = \frac{d-1}{2}$ on the hyperbolic cylinder 
$S^1_q\times AdS_d$. 
To summarise the gauge invariant partition function of conformal $p$ forms is given by the following 
determinant
\begin{eqnarray}
    \mathcal{Z}[S^1_q\times AdS_d]
    =\left[\frac{1}{\det_T(-\partial_{\tau}^2-\Delta_p)}\right]^{\frac{1}{2}}, \qquad p = \frac{d-1}{2} 
\end{eqnarray} 
Here $\Delta_p$ is the Hodge-deRham Laplacian acting on co-exact forms.  The operator $\partial_\tau^2$ 
picks out the Kaluza-Klein mass along the $S^1$ direction.

Let us follow the same analysis we carried for conformal scalars on hyperbolic cylinders. 
The eigen values of the Hodge-deRham operator acting on co-exact $p$-forms are given by 
\begin{equation} \label{eigenhdr}
\Delta_p\psi^{\{ \lambda, u\}}_{i_1i_2  \cdots i_p}= 
- \lambda^2   
 \psi^{\{ \lambda, u\}}_{i_1i_2 \cdots i_p}.
\end{equation}
Here $\{u\}$ refer to other quantum numbers on $AdS_d$ and $\psi$ refers to the eigen functions. 
Using these eigen values and the Fourier expansion on $S^1$, the partition function can 
be written as 
\begin{equation}
\log  \mathcal{Z}[S^1_q\times AdS_d] = 
-\frac{1}{2} \sum_{n=-\infty}^\infty \int d\lambda \mu_p^{(d)} (\lambda)\log \left( 
\frac{n^2}{q^2} + \lambda^2 \right) .
\end{equation}
Following the same steps as in equations (\ref{feq1}) to (\ref{geoms}) we get
\begin{equation} \label{basiccf}
\log  \mathcal{Z}[S^1_q\times AdS_d] = \frac{1}{ 2\pi i } \log\tilde R
\int_{C_o} \frac{du}{2u} \frac{ 1+ e^{ -\frac{u}{q}}}{ 1- e^{ -\frac{u}{q}}} W_p^{(d)} ( u) ,
\end{equation}
where $W_p^{(d)} (u) $ is the Fourier transform of the Plancherel measure of the 
$ p = \frac{d-1}{2}$-form in $AdS_d$. 
From (\ref{niceform}) we see that this is given by 
\begin{eqnarray} \label{conpfv}
W_p^{(d)} (u) = 
\frac{1+e^{-u}}{1-e^{-u}}\sum_{i=0}^{p}(-1)^i\binom{d-1-2i}{p-i}\frac{e^{-u(d-1-p-i)}}{(1-e^{-u})^{d-1-2i}}, 
\qquad p = \frac{d-1}{2} 
\end{eqnarray} 
Here we have replaced $d\rightarrow  d-1$ in (\ref{niceform}) and the exponential factor on the LHS 
becomes trivial since $ p = \frac{d-1}{2}$. 
Now  remarkably  the  equation in (\ref{conpfv})   can be re-written  as  
\begin{eqnarray}
W_p^{(d)} (u) &=&   \frac{1+e^{-u}}{1-e^{-u}}\sum_{i=0}^{p}(-1)^i\binom{d-1-2i}{p-i}\frac{e^{-u(d-1-p-i)}}{(1-e^{-u})^{d-1-2i}}, \\ \nonumber 
&=& \sum_{i=0}^{p}(-1)^i \binom{d}{p-i}\frac{e^{-u(p-i)}+e^{-u(d-p+i)}}{(1-e^{-u})^{d}} ,  \qquad p = \frac{d-1}{2} 
\end{eqnarray}
What is important to note is that in the second line consists of sum of 
$SO(1, d+1)$ Harish-Chandra characters of the same dimension $d$. 
Substituting this expression into (\ref{basiccf}) we obtain 
\begin{eqnarray}\label{basiccf1}
\log  \mathcal{Z}[S^1_q\times AdS_d] &=&
\frac{1}{ 2\pi i } \log\tilde R
\int_{C_o}  \frac{du}{2u} \frac{ 1+ e^{ -\frac{u}{q} } }{ 1- e^{ -\frac{u}{q}} } 
\sum_{i=0}^{p}(-1)^i \binom{d}{p-i}\frac{e^{-u(p-i)}+e^{-u(d-p+i)} }{(1-e^{-u})^{d} }, \nonumber  \\
&=&  \int_{C_o}  \frac{du}{2u}  \frac{ 1+ e^{ -\frac{u}{q}}}{ 1- e^{ -\frac{u}{q}}}  
\sum_{i = 0}^p (-1)^p \chi^{dS}_{(d, p-i)} (u) , \qquad p = \frac{d-1}{2} 
\end{eqnarray}
From the last line we see that at $q=1$ the gauge invariant partition function of the conformal $p$-form 
on $S^1_q \times AdS_{d}$ can be written as a character integral of 
only the bulk characters of the partition function of the $p$ forms on the sphere $S^{d+1}$. 
That is consider the determinant of all the co-exact forms that appear in the one loop partition function
in (\ref{gfpform}). 
Then only the bulk character of each of the  $p$-form appears in (\ref{basiccf1}), the edge modes are 
missing. 
This behaviour was observed in the study of the entanglement entropies of the Maxwell in $d+1 =4$
\cite{Casini:2013rba,Donnelly:2015hxa,Donnelly:2014fua,Huang:2014pfa,Casini:2015dsg}
field as well as the $2$ form in  in $d+1 =6$ \cite{Nian:2015xky,Dowker:2017flz,David:2020mls} and 
seen by 
Partition functions on the corresponding 
hyperbolic cylinders miss the edge modes. 
Here we see that this is true for conformal forms in all even dimensions and it can be seen at the 
level of the integrand in the character integral representation. 

Thus entanglement entropy across spherical entangling surfaces
 evaluated using partition functions on hyperbolic cylinders miss out the 
edge modes. In \cite{Soni:2016ogt,Moitra:2018lxn} it was shown that the edge modes
of the $1$-form in $d+1 =4$ dimensions  contribute only to the 
classical part of the entanglement entropy is  non-extractable. It will be interesting to repeat this analysis
 for $p$-forms and demonstrate that the edge contributions are classical and non-extractable.

\subsection{$p$-forms on branched spheres} \label{branch}

In section \ref{conscalbrsp}  we derived the character integral  representation of the one loop partition function of the 
conformal scalar on branched spheres. 
In this section we wish to generalise that computation to arbitrary $p$-forms. 
For this purpose we would need the eigen values and the corresponding degeneracies of 
the Hodge-deRham Laplacian on branched spheres. 
As far as we are aware  it is only for the $1$-form, these properties are known explicitly in 
arbitrary dimensions. Though there are generating functions for degeneracies,  from which one can 
possibly obtain the degeneracies for other $p$-forms \cite{Dowker:2017flz}. 
Therefore we will first focus on the co-exact $1$-form in arbitrary dimensions and 
obtain the character integral representation of the  one loop partition function. 
From this result  we propose the character integral for arbitrary  co-exact $p$-forms on 
branched spheres.  We will  demonstrate that the proposal agrees with 
earlier evaluations of the partition functions of $p$-forms on branched spheres.

\subsubsection* {1-form}

From \cite{DeNardo:1996kp} we see that the eigen values  of the Hodge-deRham Laplacian 
of the 1-form on the branched sphere $S^{d+1}_q$ are labelled by 
$2$ integers and are given by 
\begin{equation}
   \lambda^{(1)}_{n,m}=(n+\frac{m}{q})(n+\frac{m}{q}+d)-1 +d,  \qquad \hbox{with} \; n+m \geq 1,  \; n, m \in\{ 0, 1, 2, \cdots\}
\end{equation}
The result for the eigen values in \cite{DeNardo:1996kp}  was written for the 
vector Laplacian, but we have shifted it by  $d$
which arises due to the curvature terms relating the Hodge-deRham Laplacian  and the usual Laplacian. 
These eigen values have degeneracies 
\begin{eqnarray}\label{bsdegen}
   g^{(1)}_{n,m=0}&=&\frac{1}{n+1}   \binom{d+n-2}{n-1}
   \{(d+1) (n-4)+(d+1)^2-n+5  \},  \qquad n = 1, 2, \cdots , 
   \\ \nonumber
        g^{(1)}_{n,m}&=&2 d \binom{d+n-1}{d-1}, \qquad n = 0, 1, \cdots \qquad m =  1, 2, \cdots 
\end{eqnarray}
The partition function is therefore given by 
\begin{eqnarray}
-\frac{1}{2}\log {\rm det}_T \Delta_1^{ S_q^{d+1} }   = -\frac{1}{2} 
\sum_{n, m =0}^\infty g_{n, m }^{(1)}  \log ( \lambda_{n, m}^{(1)} ) .
\end{eqnarray}
Again using the identity (\ref{logiden})  we rewrite the partition function as
\begin{eqnarray} \label{bs1f}
-\frac{1}{2} \log {\rm det}_T \Delta_1^{ S_q^{d+1} }  =  \int_0^\infty \frac{d\tau}{2\tau} 
\sum_{n, m =0}^\infty g_{n, m}^{(1)} ( e^{-\tau \lambda_{n, m}^{(1)} } - e^{-\tau} ) .
\end{eqnarray}
Now the second term involves the sum over degeneracies. To regulate that term, we look 
at sufficiently negative $d$ and perform the sum over $m$ using  zeta function regularization. 
We obtain 
\begin{equation}
\sum_{n =1}^\infty \sum  g_{n, m=0}^{(1)}  = 1, 
\qquad\qquad
\sum_{n =0}^\infty\sum_{m =1}^\infty g_{n, m}^{(1)} =0.
\end{equation}
Therefore the equation (\ref{bs1f} )reduces to 
\begin{equation} \label{bs1f1}
-\frac{1}{2} \log {\rm det}_T \Delta_1^{ S_q^{d+1} } =    \int_0^\infty \frac{d\tau}{2\tau} 
e^{-\frac{\epsilon^2}{ 4\tau} } \left(  \sum_{n , m =0}^\infty g_{n, m }^{(1)}  - 1 \right) ,
\end{equation}
Where we have use the short distance cut off to regulate the UV divergence. 
Note that the above equation is similar to (\ref{detegi3}) with $p=1$. 
In fact the second term in (\ref{bs1f1})  can be absorbed in the first term by 
noting from (\ref{bsdegen}) that 
\begin{eqnarray} 
g_{0, m =0}^{(1)} = 0,  \qquad \qquad g_{-1, m =0}^{(1)}  = 1, \qquad \lambda_{-1, m =0}^{(1)} = 0 .
\end{eqnarray}
Therefore  we rewrite  (\ref{bs1f1})  as
\begin{equation}\label{bs1f2}
-\frac{1}{2} \log {\rm det}_T \Delta_1^{ S_q^{d+1} }   =    \int_0^\infty \frac{d\tau}{2\tau} 
e^{-\frac{\epsilon^2}{ 4\tau} }  \left( \sum_{n =-1}^\infty g_{n, m =0}^{(1)}  e^{-\lambda_{n, m =0}^{(1)} \tau}
+ \sum_{n =0, m >0}^\infty g_{n, m }^{(1)} e^{-\lambda_{n, m}^{(1)} \tau } \right).
\end{equation}
We can now follow the same steps as in (\ref{beginstep}) to (\ref{endstep}) to obtain 
\begin{eqnarray} \label{bs1f3}
-\frac{1}{2} \log {\rm det}_T \Delta_1^{ S_q^{d+1} }   =  
\int_{\epsilon}^{\infty}\frac{dt}{2 \sqrt{t^2-\epsilon^2}}(e^{( \frac{d}{2} -1) \sqrt{t^2-\epsilon^2}}+
e^{-( \frac{d}{2} -1)\sqrt{t^2-\epsilon^2}})(f_1(it)+f_2(it)), \nonumber \\
\end{eqnarray}
where 
\begin{eqnarray}
  f_1(u)&=&\sum_{n=-1}^{\infty}g^{(1)}_{n,m=0}e^{iu((n+\frac{d}{2})}
        =\frac{e^{\frac{1}{2} i (d-2) u} \left(d e^{i u}+e^{2 i u}-1\right)}{\left(1-e^{i u}\right)^d},\\
        f_2(u)&=&\sum_{m=1, n=0}^{\infty} g^{(1)}_{n,m}e^{iu((n+\frac{m}{q} +\frac{d}{2})} 
        =\frac{2 d e^{\frac{i u (d q+2)}{2 q}}}{\left(1-e^{i u}\right)^d \left(1-e^{\frac{i u}{q}}\right)}.
\end{eqnarray}
Substituting these expressions in  (\ref{bs1f3}),  rearranging the terms and taking the $\epsilon\rightarrow 0$ limit, 
one loop determinant of the co-exact 1-form on branched spheres $S^{d+1}_q$
can be written as 
\begin{eqnarray}\label{bs1f4}
-\frac{1}{2} \log {\rm det}_T \Delta_1^{ S_q^{d+1} }   &=&  
\int_0^{\infty}\frac{dt}{2t} \left\{ 
    \frac{1+e^{-\frac{t}{q} } }{1-e^{-\frac{t}{q}} } \left[ 
   \binom{d}{1}
   \frac{e^{-t}+e^{-(d-1)t} }{(1-e^{-t})^d} \right]  \right.  \\ \nonumber
   & &  \qquad\qquad\qquad 
    \left.   - \frac{1+e^{-t} }{1-e^{-t} }  \left[   \binom{d-2}{0}\frac{1+e^{-(d-2)t} }{ ( 1- e^{ -t} )^{d-2} } 
   \right]  \right\} \\ \nonumber
   &=& \int_0^{\infty}\frac{dt}{2t}   \left\{ \frac{1+e^{-\frac{t}{q} } }{1-e^{-\frac{t}{q}} } \chi_{(d, 1)}^{dS} (t) 
   -  \frac{1+e^{-t} }{1-e^{-t} }  \chi_{(d-2, 0)}^{dS} (t)  \right\}.
   \end{eqnarray}

\subsection*{Proposal for  the determinant of co-exact $p$-forms on branched spheres}

From the explicit calculation of the determinant of the $1$-form in (\ref{bs1f4}), we see that 
it is only the kinematic factor of the bulk character which acquires  dependence of the 
branching parameter$q$
for  spheres $S^{d+1}$. 
The  kinematic factor of the edge character is blind to the branching. 
Using this input, we propose that the determinant of co-exact  $p$-forms on branched spheres is given by 
\begin{eqnarray} \label{proposal}
-\frac{1}{2} \log {\rm det}_T \Delta_1^{ S_q^{d+1} }   = 
\int_0^{\infty}\frac{dt}{2t}\left\{ \frac{1+e^{-\frac{t}{q}}}{1-e^{-\frac{t}{q}}} \chi^{[dS]}_{(d,2p)}(t)
+\frac{1+e^{-t}}{1-e^{-t}} \sum_{i=1}^{p}(-1)^i\chi^{[dS]}_{(d-2i,p-i)} \right\}.
\end{eqnarray}

Using the determinant of the co-exact $p$-form on the branched sphere, one can evaluate 
the partition function of the $p$-form by using this input in the ghosts for ghosts expression 
of  (\ref{gfpform}). 
In table \ref{table5} we have listed the coefficient of the logarithmic divergence for $p$-form partition 
functions on branched spheres using the proposal (\ref{proposal}) in (\ref{gfpform}). 

As far as we are aware there are few explicit calculations of one loop determinants 
of $p$-forms on branched spheres with $p>1$.
In even $d+1$,  \cite{Dowker:2017flz} has  put forward a method to evaluate these partition functions. 
It does not rely on the explicit knowledge of the degeneracies of the Hodge-deRham Laplacian. 
As a check of the proposal in (\ref{proposal}) we have compared 
  the coefficient of the logarithmic divergence  of the partition functions in table \ref{table5}
for $p = 2$ in $d+1 =4, 6$, $p=3$ in $d+1 = 6, 8$ to that given in equation (13) of \cite{Dowker:2017flz}. 
Our values  coincides with that of \cite{Dowker:2017flz} upon identification of $q_{\rm ours} \rightarrow
\frac{1}{q_{\rm Dowker}} $. 
It is interesting to note that  $q$ deformation does not change the Hodge-duality properties of the $p$-forms. 
For instance the $(0, 2)$ pair on $S_q^4$ differ by $-2$. Similarly the $(1, 3)$ and $(2, 4)$ pair on $S_q^6$ 
differs by $2$ and $-2$ respectively. 

\begin{table}[ht]
\centering { \footnotesize{
\scriptsize{\begin{tabular}{c|l}
\hline
$d+1 $ &  ~~~~~~~~~~~~~~~~~~~~~~~ $p=0$ \\
\hline 
\\
$2$ & \quad$\frac{q^2+1}{6 q}$ 
   \\   
\\
$4$ & \quad$\frac{57 q^4+60 q^2-1}{360 q^3}$  \\
 \\
$6$ & \quad$\frac{4315 q^6+5040 q^4-245 q^2+2}{30240 q^5} $   \\
 \\
 $8$ & \quad$\frac{7 q^2 \left(33953 q^6+43200 q^4-3248 q^2+100\right)-3}{1814400 q^7} $   \\
 \\
 \\
\hline
\end{tabular}}
\quad
\vspace{5mm}
\begin{tabular}{c|l}
\hline
$d+1 $ &  ~~~~~~~~~~~~~~~~~~~~~~ $p=1$ \\
\hline 
\\
$2$ & \quad$-2$ 
\\
\\
$4$ & \quad$-\frac{33 q^4+30 q^2+1}{180 q^3}-\frac{1}{3}$ 
   \\   
\\
$6$ & \quad$\frac{2-5 q^2 \left(271 q^4+252 q^2+7\right)}{7560 q^5}-\frac{29}{90}$  \\
 \\
$8$ & \quad$-\frac{7 \left(7297 q^6+7200 q^4+98 q^2-40\right) q^2+3}{302400 q^7}-\frac{1139}{3780} $   \\
 \\
\\
\hline
\end{tabular}
\quad
\vspace{5 mm}
\begin{tabular}{c|l}
\hline
$d+1 $ &  ~~~~~~~~~~~~~~~~~~~~~~ $p=2$ \\
\hline 
\\
$4$ & \quad$\frac{57 q^4+60 q^2-1}{360 q^3}+2$ 
\\
\\
$6$ & \quad$\frac{955 q^6+840 q^4+35 q^2+2}{5040 q^5}+\frac{31}{45}$ 
   \\   
\\
$8$ & \quad$\frac{7 q^2 \left(3233 q^6+2880 q^4+112 q^2+4\right)-3}{120960 q^7}+\frac{1271}{1890}$  \\
 \\
\\
\hline
\end{tabular}
\quad
\vspace{5 mm}
\begin{tabular}{c|l}
\hline
$d+1 $ &  ~~~~~~~~~~~~~~~~~~~~~~ $p=3$ \\
\hline 
\\
$4$ & \quad$-4$ 
\\
\\
$6$ & \quad$\frac{2-5 q^2 \left(271 q^4+252 q^2+7\right)}{7560 q^5}-\frac{209}{90}$ 
   \\   
\\
$8$ & \quad$-\frac{7 \left(2497 q^6+2160 q^4+98 q^2+8\right) q^2+3}{90720 q^7}-\frac{221}{210}$  \\
 \\
\\
\hline
\end{tabular}
\quad
\vspace{5 mm}
\begin{tabular}{c|l}
\hline
$d+1 $ &  ~~~~~~~~~~~~~~~~~~~~~~ $p=4$ \\
\hline 
\\
$6$ & \quad$\frac{4315 q^6+5040 q^4-245 q^2+2}{30240 q^5}+4$ 
   \\   
\\
$8$ & \quad$\frac{7 q^2 \left(3233 q^6+2880 q^4+112 q^2+4\right)-3}{120960 q^7}+\frac{5051}{1890}$  \\
 \\
\\
\hline
\end{tabular}
\quad
\vspace{5 mm}
\begin{tabular}{c|l}
\hline
$d+1 $ &  ~~~~~~~~~~~~~~~~~~~~~~ $p=5$ \\
\hline 
\\
$6$ & \quad$-6$ 
\\
\\
$8$ & \quad$-\frac{7 \left(7297 q^6+7200 q^4+98 q^2-40\right) q^2+3}{302400 q^7}-\frac{16259}{3780}$ 
   \\   
\\
\hline
\end{tabular}}}
\caption{ Coefficient of the logarithmic divergence of the partition function of $p$-forms 
on branched spheres in even dimension using  (\ref{proposal}) } \label{table5}
\end{table}

\section{Conclusions}

In this paper we have generalized the construction given in \cite{Anninos:2020hfj,Sun:2020ame} 
of the one loop partition function  on spheres and anti-de Sitter space 
in terms of Harish-Chandra characters to $p$-forms. 
We have also seen how character integral representations make 
relations between partition functions manifest. 
For instance the equivalence of the conformal scalar partition function 
on the hyperbolic cylinder and the branched sphere was manifest 
from the character integral representations. 
It also provided insights in entanglement entropy for conformal 
$p$-forms.

In \cite{Sun:2020ame} the character integral 
representation was useful to evaluate the partition function of higher-spin Vasiliev theories.
In this context it would be important to obtain character integral representations 
of fermionic higher spin fields. 
This would enable the revisiting various one loop calculations in the literature. 
Some examples of these are one loop calculations in 
$AdS_4\times S^7$ done by \cite{Bhattacharyya:2012ye} or that done around black holes
with near horizon geometry $AdS_2\times S^2$ in \cite{Banerjee:2011jp,Sen:2011ba}. Partition functions of 
supersymmetric higher spin theories can also be evaluated. 
The character integral representation enables writing down the partition function 
once the field content and the mass spectrum of the theory is known.  Therefore
such partition functions can be obtained with considerable ease. 

The representation of the partition function of conformal $p$-forms in terms of Harish-Chandra 
characters on the hyperbolic cylinder has allowed us to show that 
partition function on hyperbolic cylinders capture the bulk contribution 
to the entanglement entropy. 
In \cite{Soni:2016ogt,Moitra:2018lxn} it was shown for the $1$-form in $4$ dimensions the contribution of
the edge modes  
to the entanglement entropy are classical and non-extractable. 
It would be interesting to   repeat this analysis for the arbitrary $p$-form. 
A similar question can be addressed for gravitons.
In \cite{David:2020mls} it was shown that the entanglement entropy of linearised gravitons
or higher spin fields 
in $d+1 =4$ 
evaluated using the hyperbolic cylinder method and that from the branched 
sphere differ by the  partition function on the sphere $S^2$ which constitute the edge modes. 
Here too it would be interesting to show that these modes are non-extractable or 
classical.

 \acknowledgments
We wish to thank Parthiv Haldar  for helping to draw the figures in  this paper. 

 \appendix
 
 \section{Harish-Chandra characters}
 
 Here we briefly list the various Harish-Chandra characters used in the paper. 
 For more details see \cite{Basile:2016aen,Anninos:2020hfj,Sun:2020ame}. 
 
 \paragraph{$AdS$ characters:}
 The easiest to understand are the $SO(2, d)$ characters associated with the 
 minimally scalars on $AdS_{d+1}$. Let the mass of the scalar in $AdS_{d+1}$ be $m$, 
 The scaling dimension or the value of the Cartan $H$  in a $SO(2, 1)$ of the lowest weight of this
  representation is given by 
 \begin{equation}
 \Delta_+  = \frac{d}{2} + \sqrt{   \frac{d^2}{4}  + m^2} .
 \end{equation} 
 Then the  Harish-Chandra character of this  massive  representation is given by 
 \begin{equation} \label{basicadsch}
 \chi_{(d, 0) m}^{AdS} (t) = {\rm Tr}_{{\cal H}_{\Delta_+}} ( e^{-Ht} )   
 = \frac{ e^{ - t\Delta_+}}{ ( 1- e^{- t})^d } .
 \end{equation}
 Here the trace is taken over  ${\cal H}_{\Delta_+}$ all in the representation. For simplicity let us take
 $t$ real and $t>0$. 
Setting $m=0$ in  (\ref{basicadsch})  we obtain
  \begin{equation}
 \chi_{(d, 0) }^{AdS} (t) 
 = \frac{ e^{ - td}}{ ( 1- e^{- t})^d } .
 \end{equation}
 We call this the the character of the $0$ form.

 Consider the  co-exact $p$-form  which satisfy the equations of motion 
 \begin{equation}
 ( \Delta_T + m^2 ) A_{i_1, \cdots i_p}  =0,
 \end{equation}
 where $\Delta_T$ is the Hodge-deRham Laplacian on $AdS_{d+1}$. The scaling dimension of the 
 lowest weight of this representation is given by  \cite{Witten:1998qj} 
 \begin{equation}\label{wavepform}
 \Delta_+ = \frac{d}{2} + \sqrt{  \left(  \frac{d}{2} - p \right)^2 + m^2 }.
 \end{equation}
 The Harish-Chandra character of this representation is given by 
 \begin{equation}
  \chi_{(d, p) m}^{AdS} (t)   = \binom{d}{ p}  \frac{ e^{ -t \Delta_+} } { ( 1- e^{-t})^d }.
  \end{equation}
Setting $m^2 =0$ in this expression we obtain
  \begin{equation}
  \chi_{(d, p) }^{AdS} (t)   = \binom{d}{ p}  \frac{ e^{ -t ( d-p) } } { ( 1- e^{-t})^d}.
  \end{equation}
   Here we would like to mention that though we have obtain the above expression 
 by setting $m=0$ in the Harish-Chandra character corresponding  to the massive representation
 it does not correspond to the  characters of the unitary irreducible 
 representations  belonging to the exceptional series.   In this paper we will loosely refer 
 to the above limit of the massive character of the $p$ form as a Harish-Chandra character. 
 This character forms the basic building blocks of all the character integral representations 
 for $p$-forms on $AdS$-spaces.

  \paragraph{$dS$ characters:}
  Let us  discuss more about the de Sitter group $SO(1, d+1)$.  
  Consider 
a minimally coupled scalar of mass 
\begin{equation}
m^2 =  \frac{d^2}{4}  + \nu^2.
\end{equation}
Then the corresponding Harish-Chandra character of this representation is given by 
\begin{equation}
\chi_{(d, 0) \nu}^{dS} (t) = \frac{ e^{ - t\Delta_+} + e^{- t\Delta_-} }{ ( 1- e^{-t} )^d}, 
\qquad \qquad \Delta_{\pm} = \frac{d}{2} \pm i \nu .
\end{equation}
Setting $m=0$ we obtain 
\begin{equation}
\chi_{(d, 0) }^{dS} (t) = \frac{ e^{ - td } + 1 }{ ( 1- e^{-t} )^d}, 
\end{equation}
As mentioned for the $AdS$ case we will still refer to the above character as 
as Harish-Chandra character though it does not correspond to a character of any UIR. 
While the conformally coupled scalar on $S^{d+1}$ has mass 
\begin{equation}
m^2_{\rm conf} =  \frac{ d^2 - 1}{4} , \qquad\qquad  i\nu_{\rm conf} = \frac{1}{2} .
\end{equation}
Therefore its character becomes
\begin{equation}
\chi_{(d, 0) {\rm conf} }^{dS} (t) = \frac{ e^{ - \frac{d-1}{2} t } + e^{ - \frac{ d+1}{ 2} t} }{ ( 1- e^{-t})^d}.
\end{equation}

Finally consider the  $p$-form with satisfies the equation (\ref{wavepform}) on $S^{d+1}$. 
Here we define $\nu$ such that 
\begin{equation}\label{nupform}
m^2 = \left(  \frac{d}{2} - p \right)^2 + \nu^2.
\end{equation}
The character of corresponding to the $p$-form is given by 
\begin{equation}  \label{masspfch}
\chi_{(d, p) \nu}^{dS} (t) = \binom{d}{ p} \frac{ e^{ - t\Delta_+} + e^{- t\Delta_-} }{ ( 1- e^{-t} )^d},
\end{equation}
where $\nu$ is obtained by solving  (\ref{nupform}).  
Setting $m=0$ in  (\ref{masspfch}) we obtain 
\begin{equation}\label{mzchr}
\chi_{(d, p) }^{dS} (t) = \binom{d}{ p} \frac{ e^{ - t( d- p) } + e^{-t p } }{ ( 1- e^{-t} )^d}.
\end{equation}
Again we wish to emphasise that in this paper we will still refer to 
to the above expression as a Harish-Chandra character for convenience  though it does not correspond 
to any character of UIR of massless $p$-forms. The building blocks of the integral 
representations for the partition functions of $p$-forms on spheres are the  characters in 
(\ref{mzchr}). 
The  reason the characters of the de Sitter group appear though we are on the sphere is due to the fact that the 
Wick rotated  de Sitter space in the static patch is a sphere.

\section{ $W_p(u)$ by direct Fourier transform} \label{contourpes}

In this appendix we evaluate the Fourier transform of the Plancherel measure $\mu_p(\lambda)$ directly 
by re-writing the transform as a sum of residues. 
This can be done for $d+1$ even. We will see that the result agrees with (\ref{niceform}) obtain 
by the differential equation method developed in the main text. 
We start with the Fourier transform.  
\begin{align}
    W_p(u)&=\int_{-\infty}^{\infty} d\lambda e^{i \lambda u} \mu_p(\lambda).
\end{align}
We can choose $u>0$ and close the contour to be the large semi-circle in the upper half plane. 
Therefore the integral reduces to the sum over residues at 
$\lambda=i(n+\frac{1}{2})$ and for $u<0$ we sum over residues at  $\lambda=-i(n+\frac{1}{2})$. 
The result is analytic in $u$. Let us choose the upper half plane to perform the integration. 
We obtain 
\begin{align}
    W_p&=
   \sum_{n=0}^{\infty} \left. {\rm Res}\left(e^{i \lambda u} {\mu}_p(\lambda)\right)
   \right|_{\lambda=i(n+\frac{1}{2})}\nonumber\\
   & =\sum_{n=0}^{\infty}-\frac{\left((2 n+1) \Gamma (d+1) e^{-\left(n+\frac{1}{2}\right) u}\right) \prod _{j=\frac{1}{2}}^{\frac{d}{2}} \left(j^2+\left(i \left(n+\frac{1}{2}\right)\right)^2\right)}{\left(\left(\frac{d}{2}-p\right)^2+\left(i \left(n+\frac{1}{2}\right)\right)^2\right) (\Gamma (p+1) \Gamma (d+1-p))}\nonumber\\
   &=e^{(\frac{d}{2}-p) u}\frac{1+e^{-u}}{1-e^{-u}}\sum_{i=0}^{p}(-1)^i\binom{d-2i}{p-i}\frac{e^{-u(d-p-i)}}{(1-e^{-u})^{d-2i}}.
\end{align}
Observe that this  result coincides with (\ref{niceform}) obtained using the differential equation method. 
It is analytic in $p, d$ and therefore can be extended to all values of $p, d$. 

\section{Massless symmetric  traceless tensors of rank spin $s$}\label{spin-s-sum}
The partition function of massless symmetric spin $s$ field on $S^{d+1}$ is given by
\cite{Giombi:2014yra}
\begin{align}
    \mathcal{Z}_s&=\left(\frac{-\Delta_{(s-1)\perp}-(s-1)(s+d-2)}{-\Delta_{(s) \perp}+s-(s-2)(s+d-2)}\right)^{\frac{1}{2}}
\end{align}
Here $\Delta_{(s)\perp}$ refers to the Laplacian on $S^{d+1}$  acting on the transverse traceless 
spin $s$ field. 
The mass term arises due to the  curvature  coupling.
The eigen-value and the degeneracy is given by
\begin{align}
    \begin{split}
    \lambda_n^{(s)}&=n(n+d)-s\\
        g_n^{(s)}&=\frac{(d+2 n) (d+2 s-2) (n-s+1) (d+n-2)! (d+s-3)! (d+n+s-1)}{(d-2)! d! (n+1)! s!}\\
    \end{split}
\end{align}
Using the same  procedure followed in section \ref{section1}, 
 the free energy can be written as
\begin{align}\label{spinspartition}
- \log\mathcal{Z}_s
  & =\int_0^\infty \frac{d\tau}{2\tau}\bigg[  \left(\sum_{n =s}^\infty
    g_{n }^{(s)} (  e^{ - \tau ( \lambda_{n}^{(s)}+s-(s-2)(s+d-2))}  - e^{-\tau} )\right)\nonumber\\
   & -\left( \sum_{n =s-1}^\infty g_{n }^{(s-1)}(  e^{ - \tau ( \lambda_{n}^{(s-1)}-(s-1)(s+d-2) )}  - e^{-\tau} )\right) \bigg].
\end{align}
The sum over the degeneracy $g_n^{(s)}$ can again be performed by working at 
sufficiently negative $d$  and analytically continuing the result to positive $d$. 
Note that the large $n$ behaviour of the degeneracy is 
\begin{align}
    g_{n }^{(s) }=\left(\frac{1}{n}\right)^{-d} \left(\frac{2 (d+2 s-4) (d+s-5)!}{(d-4)! (d-2)! s! n^2}+O\left(\left(\frac{1}{n}\right)^{5/2}\right)\right)
\end{align}
Therefore the sum $\sum_{n=s}^{\infty}  g_{n }^{(s)}$ converges in the large negative value of $d\leq -2$ .  We evaluate the sum and observe that 
\begin{align} \label{sumrelten}
    \sum_{n=s}^{\infty}  g_{n }^{(s)}
    &=\frac{(d+2 s-3) (d+2 s-2) (d+2 s-1) \Gamma (d+s-2) \Gamma (d+s-1)}{s! \Gamma (d) \Gamma (d+2) \Gamma (s)}\nonumber\\
    &=-\sum_{n=-1}^{s-1}  g_{n }^{(s)}
\end{align}
This remarkable relation allows us to  extend the sum from $n=-1$ to $n=\infty$.
Just to be explicit, 
we present the sum of degeneracy for few cases in a table \ref{table6}  
\begin{table}[ht]
\centering { \footnotesize{
\begin{tabular}{c|l|c}
\hline
$s $ & \quad $~~~~~~~~~~~~~~~~~~~~~~~g_n^{(s)} $ & $ \sum_{n=s}^{\infty}  g_{n }^{(s)}$ \\
\hline 
& &  \\
$0$ & \quad$\frac{(d+2 n) \Gamma (d+n)}{d! n!} $ & $0$
   \\   
&  & \\
$1$ & \quad$\frac{n (d+n) (d+2 n) \Gamma (d+n-1)}{\Gamma (d) \Gamma (n+2)}$ & $1$  \\
& &  \\
$2$ & \quad$\frac{(d-1) (d+2) (n-1) (d+n+1) (d+2 n) \Gamma (d+n-1)}{2 d! \Gamma (n+2)}  $ &  $\frac{1}{2} (d+2) (d+3)$  \\
&  & \\
$3 $& \quad$\frac{(d+4) (n-2) (d+n+2) (d+2 n) \Gamma (d+n-1)}{6 \Gamma (d-1) \Gamma (n+2)} $ &
$\frac{1}{12} d (d+3) (d+4) (d+5)$  \\
& &  \\
$4$& \quad $\frac{(d+1) (d+6) (n-3) (d+n+3) (d+2 n) \Gamma (d+n-1)}{24 \Gamma (d-1) \Gamma (n+2)}$
& $\frac{1}{144} d (d+1) (d+2) (d+5) (d+6) (d+7)$\\
&  & \\
\hline
\end{tabular}
\caption{ Examples illustrating $ \sum_{n=s}^{\infty}  g_{n }^{(s)}= -\sum_{n=-1}^{s-1}  g_{n }^{(s)}$  }
\label{table6}
}}
\end{table}

 Due to the relation (\ref{sumrelten}) we can re-write \eqref{spinspartition} as
\begin{align}
- \log\mathcal{Z}_s
  & =\int_0^\infty \frac{d\tau}{2\tau} \bigg[ \left(\sum_{n =-1}^\infty
    g_{n }^{(s)} (  e^{ - \tau ( \lambda_{n}^{(s)}+s-(s-2)(s+d-2))}  \right)\nonumber\\
    &-\left( \sum_{n =-1}^\infty g_{n }^{(s-1)}(  e^{ - \tau ( \lambda_{n}^{(s-1)}-(s-1)(s+d-2) )} \right) \bigg].
\end{align}
 We  now  perform the Hubbard-Stratonovich trick  given in  (\ref{contint}) and  follow the same
    steps as in equation (\ref{beginstep}) to (\ref{endstep}) and  obtain 
    \begin{eqnarray} \label{spin s simplified}
      \log\mathcal{Z}_s & = \int_{\epsilon}^{\infty}\frac{dt}{2\sqrt{t^2-\epsilon^2}}\bigg[\left(e^{i\nu_s\sqrt{t^2-\epsilon^2}}+e^{-i\nu_s\sqrt{t^2-\epsilon^2}}\right)f^{(s)} (i t)\nonumber\\
      &+\left(e^{i\nu'_{s-1}\sqrt{t^2-\epsilon^2}}+e^{-i\nu'_{s-1}\sqrt{t^2-\epsilon^2}}\right)f^{(s-1)}(it)\bigg]
    \end{eqnarray}
    with $\nu_s=i(s+\frac{d-4}{2})$ , $\nu'_{s-1}=i(s+\frac{d-2}{2})$ and 
    \begin{align}\label{spinsum}
        f^{(s)}(u)&=\sum_{n=-1}^{\infty}g^{(s)}_n e^{i u(n+\frac{d}{2}})
    \end{align}
 Substituting \eqref{spinsum} in  \eqref{spin s simplified} and taking the limit $\epsilon\rightarrow 0$ we obtain
 \begin{align} \label{parttensor}
 \log\mathcal{Z}_s & =\int_0^{\infty}\frac{dt}{2t}\bigg[\sum_{n=-1}^{\infty}g_n^{(s)}(e^{-t(n-s+2)}+e^{-t(n+s+d-2)}) \nonumber\\
 &-\sum_{n=-1}^{\infty}g_n^{(s-1)}(e^{-t(n+s+d-1)}+e^{-t(n-s+1)})\bigg]
 \end{align}
 This agrees with the (G.9) of \cite{Anninos:2020hfj}. Therefore we obtain  directly the `naive' character of massless symmetric rank $s$ tensor on $S^{d+1}$.  Note that first two terms correspond in (\ref{parttensor}) correspond to the  one loop determinant of the transverse traceless
  spin $s$ field.  It does not have any non-local terms just as in the case of the co-exact $p$-form
discussed  in section \ref{section1}.

\bibliographystyle{JHEP}
\bibliography{references} 
\end{document}